\documentclass[5p]{elsarticle}

 \usepackage{epsfig}
 
\newcommand\lsim{\mathrel{\spose{\lower 3pt\hbox{$\mathchar"218$}}
    \raise 2.0pt\hbox{$\mathchar"13C$}}}
\newcommand\gsim{\mathrel{\spose{\lower 3pt\hbox{$\mathchar"218$}}
    \raise 2.0pt\hbox{$\mathchar"13E$}}}
    
\newcommand\ngrbs{374}
\newcommand\lgrbs{351}

\usepackage{amssymb}

\journal{Nuclear Physics B}

\begin{document}

\begin{frontmatter}



\title{Galaxies as seen through the most Energetic Explosions in the Universe}


\author[label1,label2]{Sandra Savaglio}

\address[label1]{Physics Department, University of Calabria, I-87036 Arcavacata di Rende, Italy}
\address[label2]{European Southern Observatory, D-85748 Garching bei M\"unchen, Germany}

\begin{abstract}
A gamma-ray burst (GRB) is a strong and fast gamma-ray emission from the explosion of stellar systems (massive stars or coalescing binary compact stellar remnants), happening at any possible redshift, and detected by space missions. Although GRBs are the most energetic events after the Big Bang, systematic search (started after the first localization in 1997) led to only \ngrbs\ spectroscopic redshift measurements. For less than half, the host galaxy is detected and studied in some detail. Despite the small number of known hosts, their impact on our understanding of galaxy formation and evolution is immense. These galaxies offer the opportunity to explore regions which are observationally hostile, due to the presence of gas and dust, or the large distances reached. The typical long-duration GRB host galaxy at low redshift is small, star-forming and metal poor, whereas, at intermediate redshift, many hosts are  massive, dusty and chemically evolved. Going even farther in the past of the Universe, at $z > 5$, long-GRB hosts have never been identified, even with the deepest NIR space observations, meaning that these galaxies are very small (stellar mass $< 10^7$ M$_\odot$). We considered the possibility that some high-$z$ GRBs occurred in primordial globular clusters, systems that evolved drastically since the beginning, but would have back then the characteristics necessary to host a GRB. At that time, the fraction of stellar mass contained in proto globular clusters might have been orders of magnitude higher than today. Plus, these objects contained in the past many massive fast rotating binary systems, which are also regarded as a favorable situation for GRBs. The common factor for all long GRBs at any redshift is the stellar progenitor: it is a very massive rare/short-lived star, present in young regions, whose redshift evolution is closely related to the star-formation history of the Universe. Therefore, it is possible that GRB hosts, from the early Universe until today, do not belong to only one galaxy population.
\end{abstract}


\end{frontmatter}


\section{Introduction}

The discovery of a galaxy hosting a gamma-ray burst (GRB) was achieved for the first time in February 1997, with the identification of an afterglow, the event GRB\,970228 (Costa et al\,1997). However, its redshift, $z=0.695$, was measured later from the detection of emission lines in the host through Keck spectroscopy (Djorgovski, et al.\,1999). The very first gamma-ray burst redshift was measured in May 1997, for GRB\,970508. The absorption lines  seen in the Keck spectrum of the optical afterglow gave $z =0.835$ (Metzger et al.\,1997).  Today, precise localization of afterglows (2 arcsecs or better) is routinely performed, at the level of several events per week, mainly thanks to the data collected with the most successful dedicated space mission, the NASA  satellite {\it Swift}, launched at the end of 2004 (Gehrels et al.\,2004). Since 1997, and as of the first half of 2015, gamma-ray instruments identified a total of more than 1,400 GRBs.  
Although $\sim 90$\% of {\it Swift} GRBs are localized, thanks to the X-ray instrument XRT, the afterglow localization for the whole population since 1997 was possible for only about half them. These are mainly long GRBs (more than 90\%), those for which the gamma-ray emission is longer than a couple of seconds, and associated with the death of a massive star (mass of the progenitor $M>30$ M$_\odot$). The hosts are  detected mainly for the long GRBs.

\begin{figure*}\label{f1}
\epsfysize=7cm
\centerline{\epsfbox{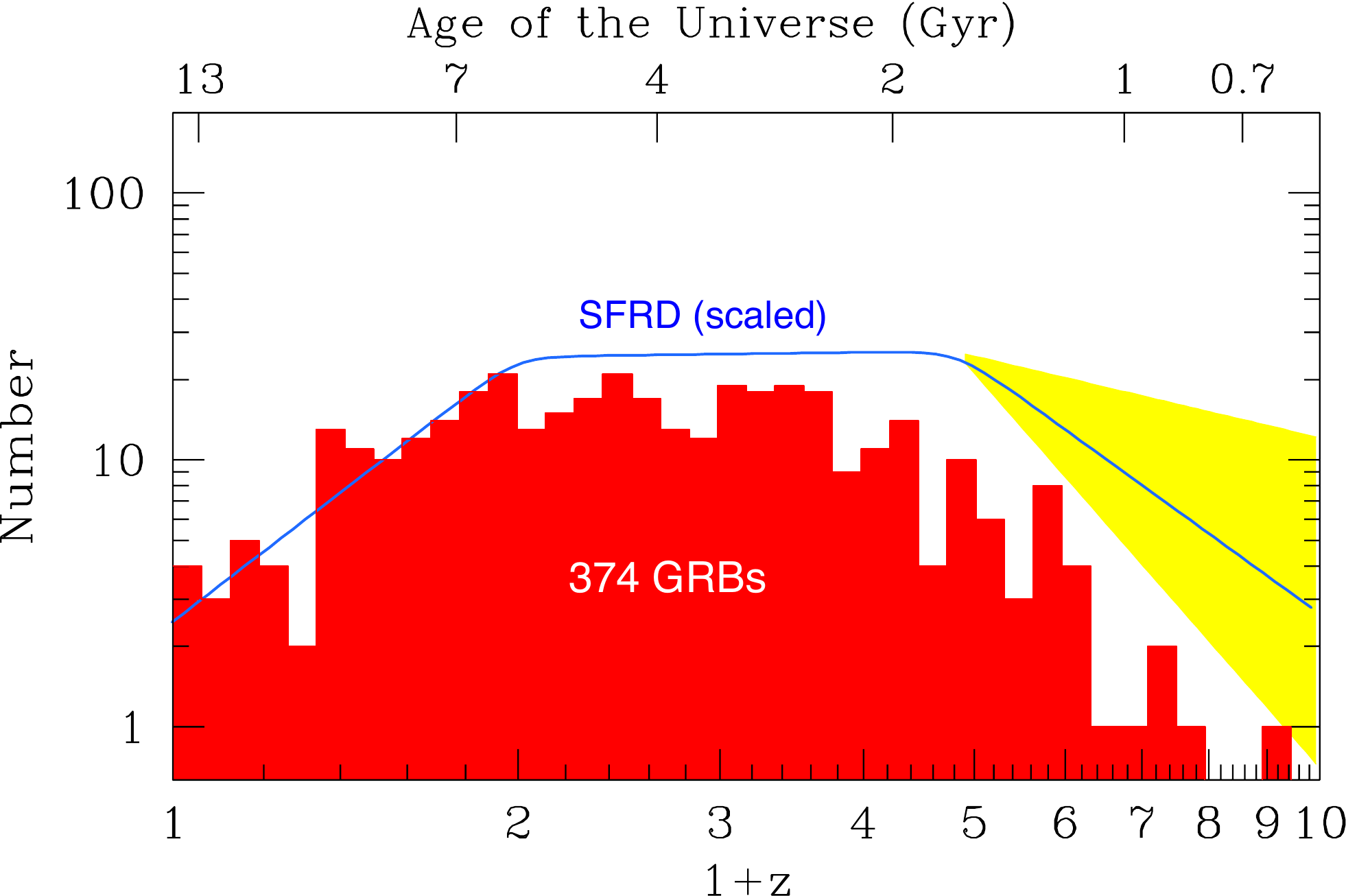}}
\caption{Number of GRBs per redshift bin (as of May 2015). All \ngrbs\ redshifts are spectroscopically determined, either from the afterglow or from the host galaxy. The blue curve and yellow shaded area represent the star-formation rate density (SFRD) of the Universe (from Wei et al.\,2014), scaled to the GRB histogram for $1+z \leq 2$. The SFRD below $1+z \sim 5$ is the one determined by Hopkins \& Beacom (2006) and Li\,(2008) from an observational compilation of UV galaxies. The $1+z>5$ SFRD (yellow shaded area) is vaguely constrained by the GRB detection rate (Chary et al.\, 2007; Y\"uksel et al.\,2008; Wang \& Dai 2009).}
\end{figure*}

Fruchter et al.\,(2006) investigated the location where the GRB takes place and found that those at $z<1.2$ prefer the most active regions in the galaxy, more than what done by core-collapse supernovae (CC-SNe). Following this early result, Kelly et al.\,(2008) found that GRB environments are more similar to those of SN type Ic rather than SN type II. This is consistent with the fact that spectroscopically confirmed SNe found a couple of weeks after the GRB are type Ic (see Hjorth \& Bloom 2012; for a review). SN Ic's tend to be more luminous than the typical CC-SNe. Studies of the SN-GRB connection is limited basically to $z< 1$, where SN can spectroscopically be identified.

At higher redshift, it has been finally found that the role of dust is important. The TOUGH sample includes a complete investigation of 69 GRB hosts (median redshift $z=2.14\pm0.18$). Hjorth et al.\,(20112) indicate that optically dark GRBs tend to occur in more massive and redder galaxies with respect to optically detected GRBs. The presence of large dust content was found in dark GRBs though radio and deep observations. The large sample SHOALS studied by Perley et al.\,(2015a,b) with multi-band optical-IR observations of 119 hosts revealed a relatively large fraction (20\%) of dust obscured galaxies, which are also massive systems. Radio observations (Micha{\l}owski et al.\,2012; Perley \& Perley 2013) have revealed that some have properties similar to sub-millimeter galaxies, but the majority does not have hidden high dust obscured star-formation rate (SFR).

Despite the numerous surveys which used with different means and selection criteria, we are still  not sure how well GRB hosts represent the whole galaxy population. For instance, contradictory results are found on the galaxy luminosity function. Using the THOUGH sample, Schulze et al.\,(2015) compared the GRB-host luminosity function to the one of Lyman break galaxies (LBGs) and concluded that GRBs select metal poor galaxies. This was not confirmed by the high-redshift $3 < z < 5$ sample studied by Greiner et al.\,(2015).

Since the beginning, it was  pursued the idea that GRBs prefer low-metallicity environments (e.g., Graham \& Fruchter 2013; Vergani et al\,2014; and references therein).  From the theoretical and modelling point of view, this conclusion was reached at low (Niino et al.\,2011; Boissier et al.\,2013; Vergani et al.\,2014) and high redshift (Chisari et al.\,2010; Trenti et al.\,2015). However, see the work by Campisi et al.\,(2011) and Elliott et al.\,(2012) for a different conclusion.

This shows that we are still far from a full comprehension and interpretation of the galaxy population hosting GRBs. Contrary to what is commonly believed, our main limitation is not the observational bias, but the small number statistics. Since the first afterglow in 1997, the number of detected hosts with spectroscopic redshift is only 1/4 (\ngrbs) of the total number of identified afterglows\footnote{However, we notice that the fraction of detected hosts is much higher and at least 80\% for well defined complete samples (e.g., TOUGH, or SHOALS surveys).}. The majority are long GRBs (\lgrbs), 23 are short\footnote{This large long-to-short number ratio is partly and likely due to the rarity of short events, but mainly determined by the faintness of their X-ray/optical afterglow, which, together with the short duration, makes a precise localization and identification more difficult}. Eighteen years after the first afterglow localization, we are still dealing with a small number statistics. Nevertheless, the fact that GRBs are distributed over the entire redshift interval ever explored in the history of human kind (Fig.\,\ref{f1}), from $z=0.0085$ (GRB\,980425 at 37 Mpc from the Milky Way; Galama et al.\,1998) to $z=8.23$ (GRB\,090423; Salvaterra et al.\,2009; Tanvir et al.\,2009a), makes them the most valuable resources of  exploration of the dawn of the Universe.

\section{The typical GRB host galaxy}

The definition of the population of galaxies hosting GRBs is not precisely defined. When an optical afterglow is detected and a spectrum is obtained, it is often possible to measure the redshift from the identification of absorption lines, for instance the strong MgII$\lambda\lambda2796,2803$ doublet (in the optical for the wide redshift interval $0.35 < z < 2.5$). We automatically assume that these absorption features are associated with the interstellar medium (ISM) of the host galaxy, thus implicitly conclude that the host is identified.  However, when we talk about the properties of the host galaxy population as a whole, we generally refer to the detected stellar and gas emission of the galaxy which lies closest to the afterglow. By doing this, we not always give the proper emphasis to the results obtained from absorption line studies, because the direct identification of the galaxy is not always possible. Our view becomes more incomplete at  $z>1.5$, where galaxies become increasingly fainter. As a consequence, the 'GRB host population' we generally discuss includes basically galaxies detected in emission, photometrically and spectroscopically, mainly at $z<1.5$, spanning about 2/3 of the entire history of the Universe. What about the GRB hosts during the first 4.3 Gyr of life?

\begin{figure}\label{f2}
\epsfysize=8.5cm
\centerline{\epsfbox{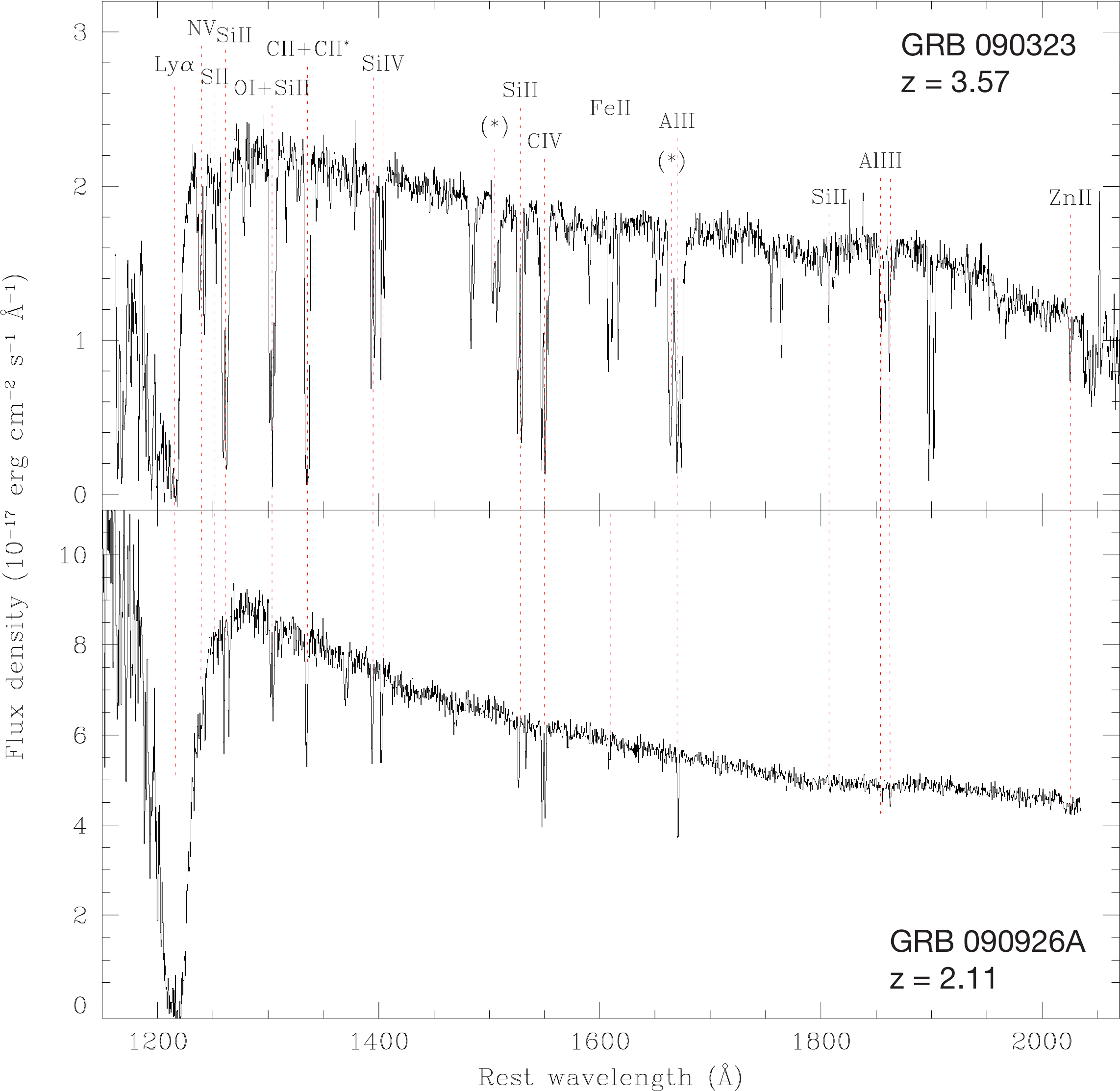}}
\caption{Comparison between two high-$z$ GRB afterglow spectra (probing the cold ISM in the host galaxy). {\it Upper panel}: spectrum of the afterglow of GRB 090323 at $z = 3.57$ (Savaglio et al.\, 2012), with strong metal absorbers revealing the presence two systems separated by 660 km s$^{-1}$ (total $N_{\rm HI}=10^{20.76\pm0.08}$) with super solar metallicity ([Zn/H]\,$= +0.25\pm0.09$). {\it Lower panel}: spectrum of the afterglow of GRB 090926 at $z = 2.1062$ (Rau et al.\, 2010), characterized by a strong HI absorption ($N_{\rm HI} = 10^{21.73\pm0.07}$) and weak metal lines, from which a low metallicity is derived ([Si/H]\,$= -1.89\pm0.17$). At $z=3.57$, the  Universe was 1.8 Gyr old and  at $z=2.11$, it aged by 1.36 Gyr.}
\end{figure}

For only less than half of all GRBs with measured redshift (160) imaging has shown the presence of a galaxy. Multi-band photometry from optical to the near infrared (NIR) at low redshift  is sufficient to derive with good accuracy the stellar mass of the galaxy. Of this population, another half or so are those with one (or more) emission lines detected, typically the strong [OII]$\lambda3727$ feature, more easily detected (in the optical/NIR) when $z < 1$. Emission lines provide a direct measurement of SFR, the most known parameter. If, together with the oxygen, one Balmer line is detected, then the metallicity can be estimated. This can be H$\beta$ for $z<1$, otherwise H$\alpha$ for $z<0.5$. At the moment, for at most 30 galaxies it was possible to determine the metallicity. When H$\alpha$ and H$\beta$ are detected together, the line ratio (Balmer decrement) gives the dust extinction in the optical $A_V$ (about 22 hosts). These samples are small, but we can improve the statistics by considering gas properties from absorption lines.

At high redshift, GRB afterglow spectroscopy provides independent measurements of the metallicity (at least 26 GRB hosts). This is possible exclusively for $z > 2$, when the Lyman-$\alpha$ absorption line (necessary to measure the HI column density) is redshifted to the optical window. This redshift limit is set by the fact that the only suitable UV spectrograph, on Hubble Space Telescope, is not used to observe GRB afterglows. If HI is not detected, low-ionization lines, associated with heavy elements in the neutral ISM, are used to derive relative abundances. For instance, the zinc-to-iron or silicon-to-iron ratio are directly connected to the amount of dust depletion, therefore, to the dust column distributed along the sight line in the host.

The overlap between these different samples is still very limited. For only a very few individual GRBs it was possible to detect the afterglow and study absorption lines, together with the host in imaging, and also perform the  spectroscopy for the emission lines, therefore, derive in principle the maximum number of galaxy parameters. However, soon the heavy use of the Optical-NIR spectrograph X-Shooter (at the Very Large Telescope) for GRB programs will deliver first results on relatively large numbers\footnote{At the time of writing, the results of a large sample of X-Shooter spectra of GRB hosts at $0.1 < z < 3.6$ are in the process of being published (Kr{\"u}hler et al.\,2015)}, which will allow the exploration of an unknown territory, in a way that even normal galaxy surveys cannot do. We have to emphasize that the scientific community dealing with GRBs cannot really be accused of being lazy for not publishing data and  results. An interesting statistics is the number of papers where the information on GRB host galaxies is reported\footnote{See the GHostS database at http://www.grbhosts.org for more information}: more or less 450 are the papers from which information for about 250 GRB hosts are reported. Yet, our understanding of this galaxy population is effected by the small number statistics. The difference with normal galaxy surveys, that count many thousands of galaxies, is apparent. For instance, the Sloan Digital Sky Survey (SDSS), 15 years after the beginning of the project includes photometry and spectroscopy for a sample of half a billion galaxies.

\begin{figure*}\label{f3}
\epsfysize=7cm
\centerline{\epsfbox{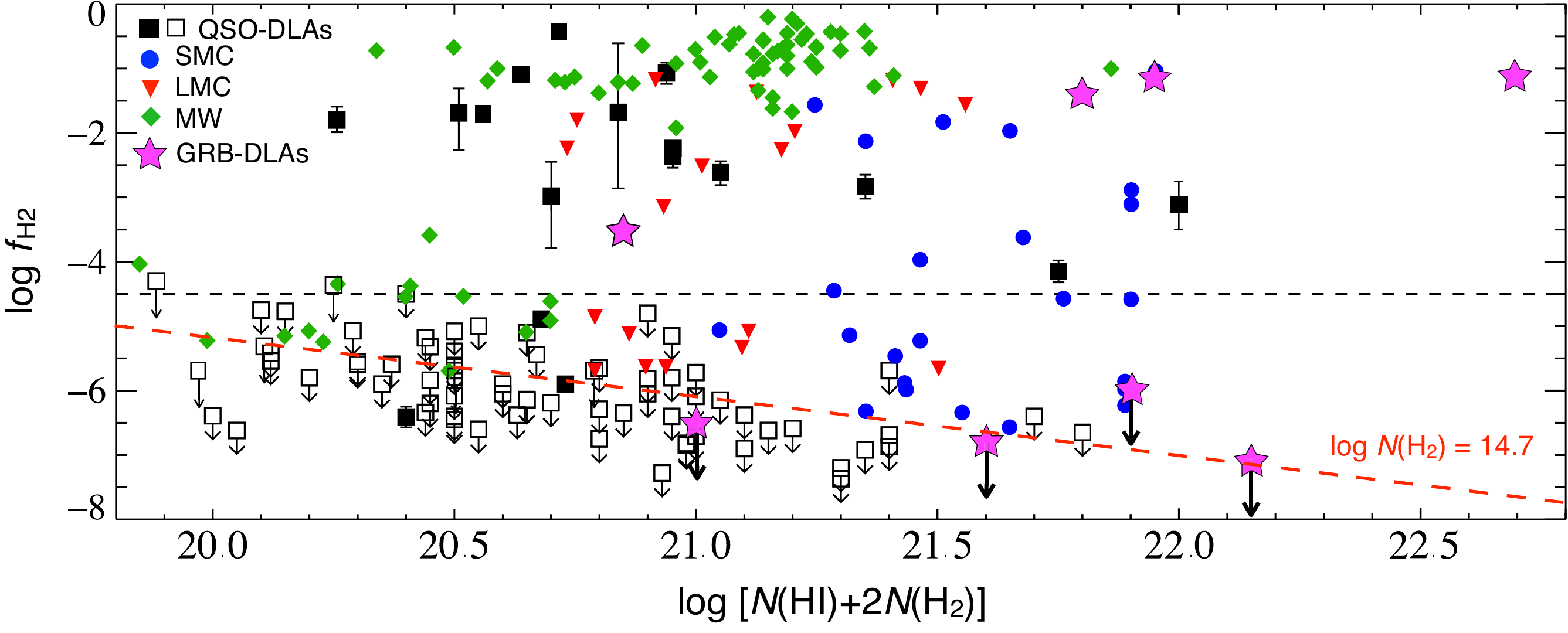}}
\caption{Fraction of column density of molecular hydrogen $f_{\rm H_2} = 2N({\rm H_2})/[N({\rm H I})+2N({\rm H_2})]$ vs.\,the total column density of hydrogen $N({\rm H I})+2N({\rm H_2})$. The blues dots, red triangles and green diamonds correspond to H$_2$ detections in the Small Magellanic Cloud (SMC, Tumlinson et al.\,2002), Large Magellanic Cloud (LMC, Tumlinson et al.\,2002), and Milky Way (MW, Savage et al.\,1977), respectively. Black filled and empty squares are measurements and upper limits in QSO-DLAs, respectively (Noterdaeme et al.\, 2008; Guimar\~aes et al.\,2012; Muzahid et al.\,2015). Pink filled stars are measurements and upper limits in GRB-DLAs (Fynbo et al.\,2006; Tumlinson et al.\,2007; Prochaska et al.\,2009; Kr\"uhler et al.\,2013; Friis et al.\,2014). The region below the dashed-red line indicates a column density of molecular hydrogen below $\log N({\rm H_2}) = 14.7$.}
\end{figure*}

For a long time and still today, the scientific community has adopted the idea that the typical GRB host galaxy is a small star-forming metal poor galaxy with low dust content (e.g., Levesque et al.\,2010a; Graham \& Fruchter 2013; Vergani et al.\,2014). For instance, a typical host is the one of GRB\,011121 (Bloom et al.\,2002; K{\"u}pc{\"u} Yolda{\c s} et al.\,2007), among the first ever studied in detail. The galaxy is at low-redshift ($z=0.362$), small (stellar mass $M_\ast = 10^{9.8\pm0.2}$ M$_\odot$), with low metallicity $\log (Z/Z_\odot) = -1.16$ and low dust extinction ($A_V=0.4$), but, given its mass, a relatively high SFR\,$\sim 2$ M$_\odot$ yr$^{-1}$. The initial view, according to which the host of a GRB, whose progenitor is a  massive and metal poor star, is special and does not represent the general galaxy population, has been hard to change. 

These common properties, derived at low redshift, can naturally be explained if we consider that galaxies are easier to observe and study if they are close, at $z<1$, where  faint targets can be detected, emission lines measured in the optical, and dust extinction, SFR and metallicity estimated.  At this low redshift, the most common star forming galaxy is small. High star formation activity is the necessary ingredient to have a high chance of a  massive-star explosion. Small galaxies are typically metal and dust poor. The natural consequence is that the host galaxy of a low-$z$ GRB is a small galaxy which is also metal and dust poor. However, extrapolating this idea to any redshift is itself a bias. Instead, it should be more natural to think that GRBs at high redshift happen in galaxies different from does at low redshift, because the Universe back then was very different. 

It would take too long to describe the large variety of properties seen lately in GRB host galaxies. We report a few examples. For instance, the two cases seen in Fig.\,\ref{f2}, where the two afterglow spectra for GRB\,090926A at $z=2.11$ (Rau et al.\,2010) and GRB\,090323 at $z=3.57$ (Savaglio et al.\,2012) are compared. Their metallicity are at two extremes: the former is about 1/100  the solar value (large HI absorption and the weak metal lines), the latter is over solar (strong metal lines and relatively low HI absorption). It is remarkable in the latter that a high chemical enrichment is measured in a galaxy when the Universe was less than 2 Gyr old. Unfortunately the host is detected, but its mass is not known.

\begin{figure*}\label{f4}
\epsfysize=8cm
\centerline{\epsfbox{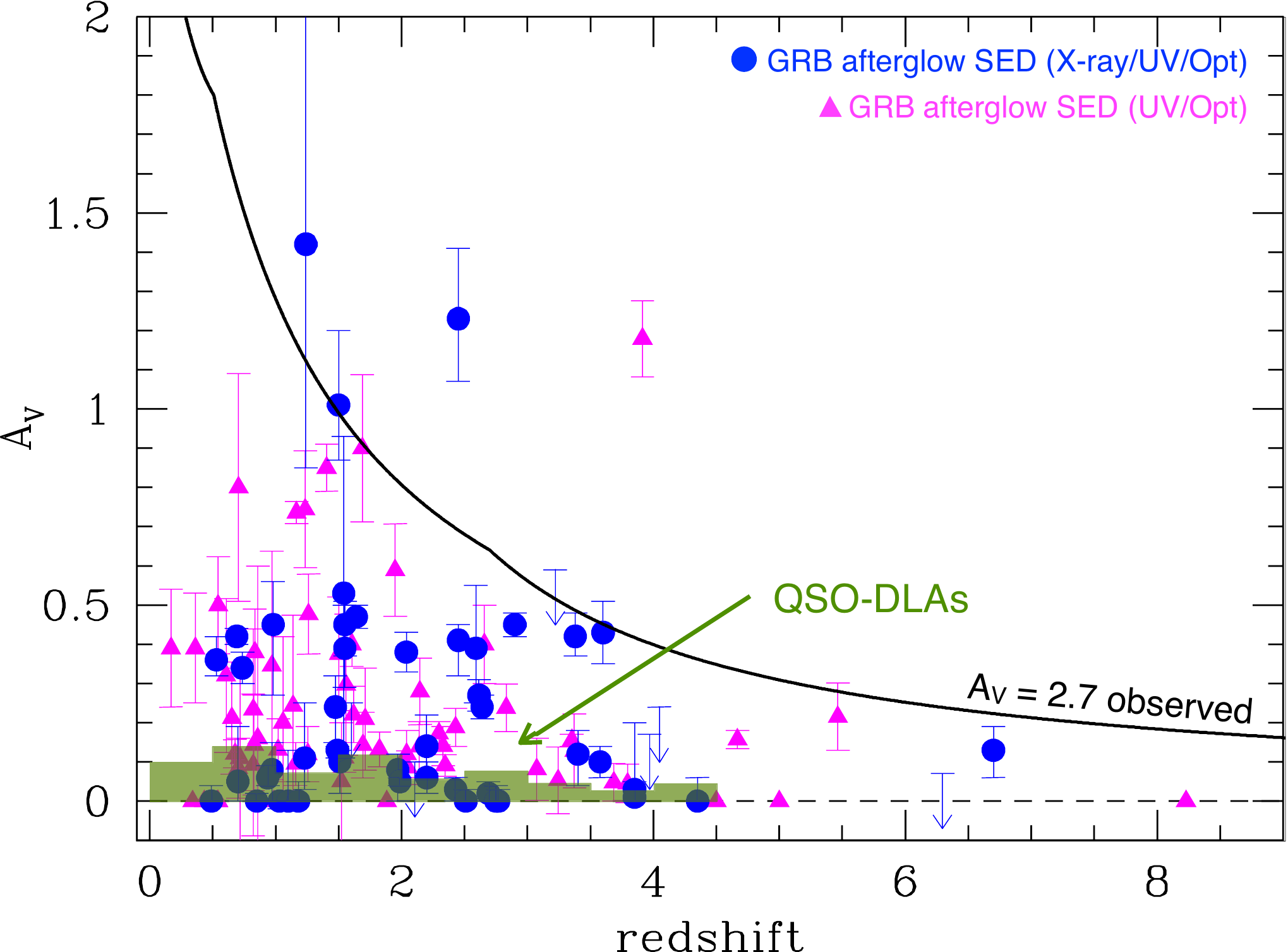}}
\caption{Dust extinction in the visual band $A_V$ derived from GRB afterglows and QSO-DLAs. Blue circles are $A_V$ derived from the SED fitting of GRB afterglows from X-ray to UV and optical (Greiner et al.\,2011; Schady et al\,2011; Zafar et al.\,2012). Filled triangles are the same, but using UV to  optical only (Kann et al.\,2006; 2010; 2011). The green histogram indicates the mean  $A_V$  and dispersion per redshift bin from a large sample of QSO-DLA, derived by using the zinc-to-iron relative abundances and assuming that these are proportional to the visual dust extinction $A_V$ (see Savaglio 2006). The black curve indicates the extinction in the rest-frame $V$-band for a constant $V$-band extinction in the observed frame $A_V=2.7$.}
\end{figure*}

\section{The meaning of the observational bias}

Since the first GRB redshift measurement in 1997, many scientists inside and out the GRB community have learned that galaxies hosting GRBs do not represent the whole galaxy population. It is  believed that GRBs are special and rare events and their hosts are observational biased. Moreover, theoretical models predict that, for a GRB to occur, a jet of relativistic particles has to break from the star collapsing core. The gamma-ray radiation has to emerge from inside and stay collimated, because if the emission were isotropic, it would be too large (equivalent to the rest-mass energy of the star). The collimation is preserved by the high angular momentum, which is possible if radiation pressure (less efficient in a low-metallicity star) is low, and the mass loss as well. Therefore, it is easier to imagine that the GRB ambient is chemically pristine.

However, it is strange that the dozen spectroscopically identified SNe, seen within 2 weeks after the GRB (mainly $z<0.5$), are type Ic: no hydrogen nor helium, but silicon. These features are explained by mass loss of the external envelope, through stellar wind (like in Wolfe-Rayet stars) or by the presence of a companion in a binary system.

From one side, this issue is fundamental and still highly debated. From the other, it was clear since the beginning of the first redshift that there was something  special about the host galaxies. They looked different, although not necessarily unique, from the observational point of view. It looked like we had missed these galaxies while using the traditional approach: observational campaigns start from optically bright targets. With GRBs, we can in principle overcome this limitation: GRBs are bright in the gamma-ray and short, and pinpoint to locations in galaxies regardless of their brightness. It is hard to imagine a less biased way of selecting galaxies. The main problem of this population right now is, as  said above, the small number statistics, and not the way (still vaguely understood) GRBs select galaxies.

The GRB host population suffers of some kind of bias, but not more than any other galaxy population we can think of. Any galaxy population is detected with specific tools and using some selection criteria, and by definition any survey is biased. Observational biases are corrected by using well tested and complex analysis methods, but this is done by assuming that we have full control of the observational tools (possible) and have an idea (up front) of what we are missing. GRB hosts are selected in a very different way, and this, by itself, can only be good because it tackles the problem of galaxy search from a different point of view. Being different is not a good reason for discarding them when studying galaxy formation and evolution. This would introduce a bias, and there is no scientific justification to do so.

GRB host galaxies have an important value for our understanding of galaxies at any epoch. In the 450 papers mentioned above, a large variety of telescopes, wavelength bands, instruments, depths and analysis methods are used. Thus, very different are the properties that we found, which is  by itself an indication that, given the small number statistics,  there cannot be a typical GRB host galaxy. The most  common comments the GRB community faces when applying for telescope time is that "the sample is too small" or "biased and does not represent the galaxy population". This is not always a good justification: we have achieved many fundamental discoveries from a small sample, adding one or a few more objects is important in any case.

\begin{figure*}\label{f5}
\epsfysize=9.9cm
\centerline{\epsfbox{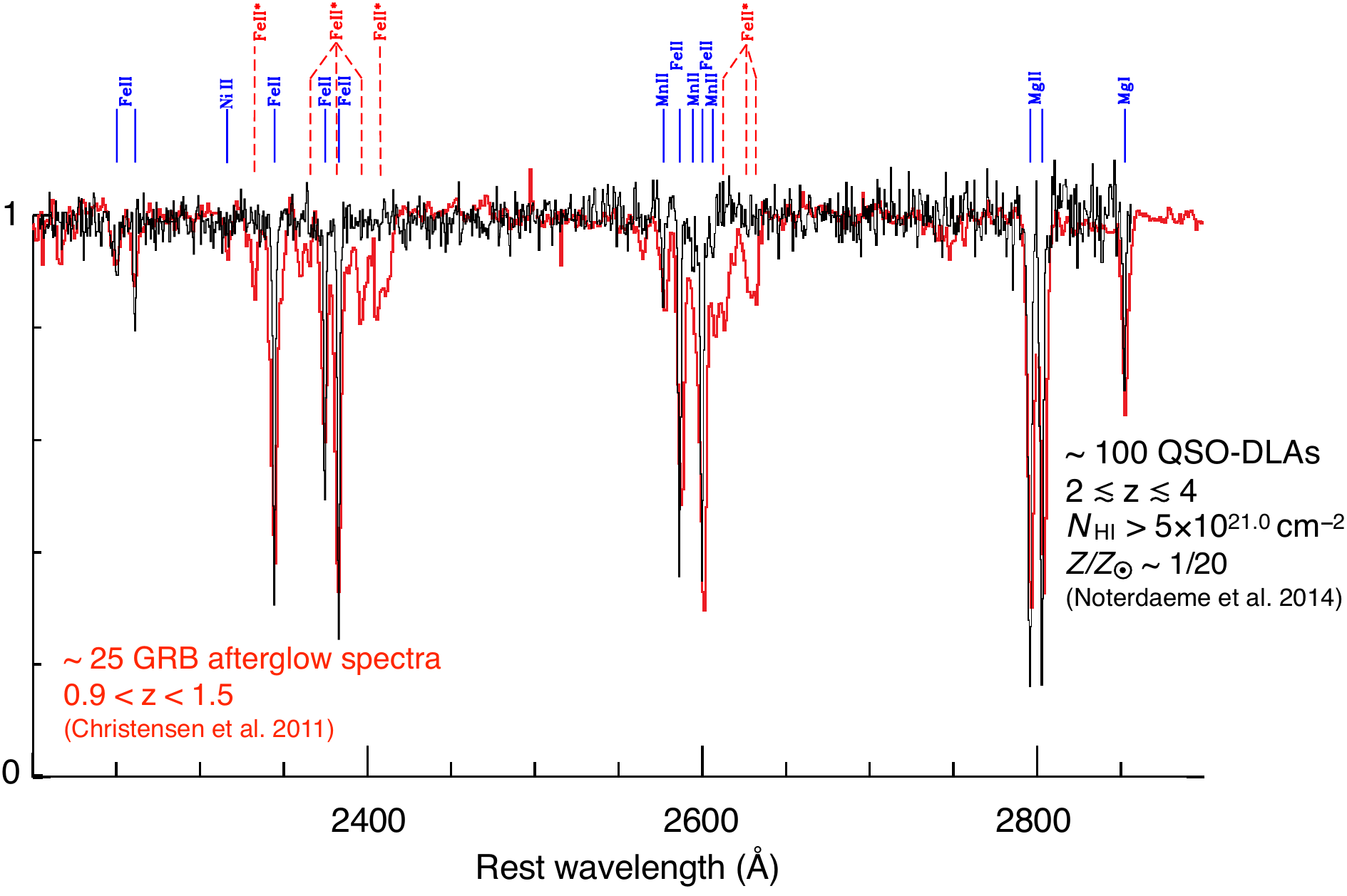}}
\caption{Comparison of two composite UV spectra probing the cold ISM in galaxies. The red spectrum is obtained from $\sim 25$ GRB afterglow spectra in the redshift interval $z = 0.9 - 1.5$ (Christensen et al. 2011). The black one includes about 100 QSO-DLA spectra in the redshift range $2\leq z \leq 4$ and is characterized by large HI column densities $N_{\rm HI} > 5\times 10^{21.0}$ cm$^{-2}$ and relatively low metallicity $Z/Z_{\odot} \sim 1/20$ (Noterdaeme et al.\,2014).  Vertical marks indicate resonance (solid blue) and fluorescent (dashed red) lines. The GRB-DLA composite is characterized by much stronger metal lines than the QSO-DLA one.}
\end{figure*}

\section{Spanning the entire history of the Universe}\label{s4}

There cannot be a typical GRB host simply because GRBs are seen from $z=0$ to the highest ever measured redshift $z=8.2$, when the first collapsed objects  formed. Over this time, the Universe has dramatically changed. Not surprising if GRB hosts did as well. It is widely accepted that the Universe was much more active in the past, as clearly seen from the redshift evolution of the star-formation rate density (SFRD). Fig.\,\ref{f1} shows the history of the SFRD, characterized by a beginning, a steady/intermediate state, and a decaying era. Our knowledge is much more accurate for the last 8 Gyr ($z< 1$), when the SFRD dropped by at least a factor of 10. This is the Universe where half of all GRB hosts are detected and from where most of our knowledge comes from.

At higher redshift, hosts are harder to see in emission because they are fainter, but  the GRB can overcome this limitation. Afterglow spectra reveal absorption lines associated with the ISM in the host.  Column densities of the gas are often large, and the absorbers are classified as damped Lyman-$\alpha$ systems (DLAs). In these cases the gas is mainly neutral, the ionization correction can be neglected and the metallicity can easily be derived. This was done for GRBs in the interval $2 < z < 6.3$. At $z> 4$, the redshift is measured for 27 GRBs from the afterglow (Fig.\,\ref{f1}), but not much luck for the direct detection of the host: only 4 have been seen, in the rest-frame UV, and none at $z\geq5$ (where 10 are the total number of GRB redshifts).

GRB-DLAs have often a large column density of HI. Of the 62 GRB-DLAs with a measured $N_{\rm HI}\geq 2\times 10^{20}$\,cm$^{-2}$, 53 (85.5\%) have $N_{\rm HI}\geq 10^{21.5}$\,cm$^{-2}$, and 13 (21\%) have $N_{\rm HI}\geq 10^{22}$\,cm$^{-2}$. The largest value ever measured ($N_{\rm HI}= 10^{22.70\pm0.15}$\,cm$^{-2}$) was found in GRB\,080607 at $z=3.0363$ (Prochaska et al.\,2009). For comparison, the fraction of DLAs with $N_{\rm HI}\geq 10^{21.5}$\,cm$^{-2}$ found along QSO sight lines is 5.4\% only. This is one of the most striking differences between what has been known for decades about the ISM in high-$z$ galaxies and what is now observed with GRB spectroscopy. 

\begin{figure*}\label{f6}
\epsfysize=6cm
\centerline{\epsfbox{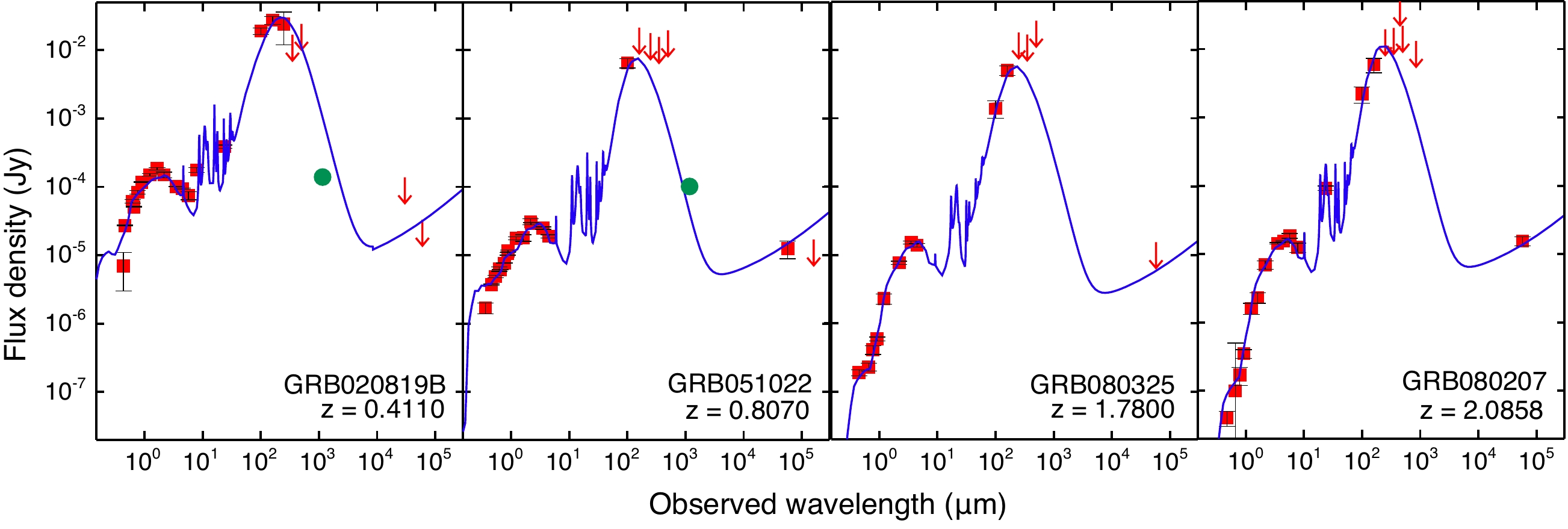}}
\caption{Examples of host galaxies associated with dark GRBs, and detected with Herschel (Hunt et al.\,2014). Filled green circles are the first ALMA detections of GRB hosts (Hatsukade et al.\,2014).}
\end{figure*}

Another science topic where GRB investigations revealed interesting results is the molecular gas. The low fraction of molecular hydrogen in QSO-DLAs is a well known phenomenon. Fig.\,3 shows the H$_2$ column densities measured in local environments (the Milky Way and Magellanic Clouds, Savage et al.\,1977; Tumlinson et al.\,2002), and in QSO-DLAs. Many are the low values and upper limits in the former population. The investigation of molecular hydrogen was possible only though a very limited number of GRB afterglows, eight in total, and half of them led to a positive detection. Remarkable is the relatively high fraction of molecular hydrogen $\log f_{\rm H_2}  = 2N({\rm H_2})/[N({\rm H I})+2N({\rm H_2})] \sim -1.14$ found in GRB\,080607 at $z=3.0363$ with the  extreme column density of neutral hydrogen $N_{\rm HI}= 10^{22.7}$ (Prochaska et al.\,2009).

With a better statistics, available in the last few years, many are the GRB hosts revealing unexpected (according to the common belief) properties. A few have high molecular gas content (see above), others are massive, many have a large amount of dust and dust extinction. The dust extinction is easily estimated by assuming that the intrinsic emission is  a simple or broken power law, and fitting the observed SED of the afterglow in NIR, optical, UV and X-ray. In Fig.\,4, the optical extinctions $A_V$ in GRB afterglows is shown together with the estimated average (plus dispersion) $A_V$ in QSO-DLAs. In the latter, $A_V$ is estimated by considering that the dust depletion, represented by the zinc-to-iron relative abundance, gives the dust content along the sight line, thus, the dust extinction (Savaglio 2006). The figure shows that GRBs probe dustier environments than QSO-DLAs.

The presence of dust is also manifested by the UV bump, the broad absorption feature centered at $\lambda \sim 2175$ \AA. Prominent in the Milky Way, it is much weaker in the LMC and absent in the SMC. Its origin is still not totally known: graphites and silicates, mixed with complex molecules? The UV bump is not easily detected outside local environments. In QSO-DLAs, it is generally very weak and can be seen only from composite spectra (see Ledoux et al\,2015 for a more detailed discussion). 

A strong UV bump is seen in several GRB afterglow spectra. For instance,  Prochaska et al\,(2009) detected it in GRB\,080607 at $z=3.036$, from which $A_V = 3.2$ was derived. In the same spectrum, a large HI column density and molecular absorption is measured as well (see Section \ref{s4}). The host of this dark GRB was widely investigated from the UV to the infrared, and a relatively high stellar mass is derived, $M_\ast \simeq 4\times10^{10}$ M$_\odot$ (Perley et al.\,2013). This is another host which, according to the canonical view, can be considered 'abnormal'.  More recently, an unusually strong UV bump is identified in GRB\,140506 at $z = 0.889$. The best-fit model gives an extreme $A_V$ (Fynbo et al.\,2014). In general, the presence of the UV bump is also demonstrated from multi-band photometry and SED fitting of the afterglow. One clear detection is in GRB\,070802 at $z=2.45$, giving $A_V = 1.8$ and a best-fit LMC extinction law (Kr{\"u}hler et al.\,2008).

\section{The properties of GRB host galaxies}

Not all GRB hosts are metal poor. At  $z <2$, a few  exceptions were found (Levesque et al.\ 2010b; Perley et al.\ 2012;  Kr{\"u}hler et al.\ 2012; Niino et al.\ 2012). At $z>2$, GRB-DLAs display a large dispersion. We already mentioned the two extreme measurements in Fig.\,\ref{f2}. The super-solar metallicity found in GRB\,090323 at $z=3.57$ (Savaglio et al.\ 2012) demonstrates that even at high redshift, the host is not necessarily metal poor.

The typical high metallicity in GRB-DLAs is more clearly seen in Fig.\,5. This displays an instructive composite spectrum in the rest-frame interval  $\lambda\lambda = 2200-2900$ \AA\ (around the FeII and MgII absorptions) from $\sim25$ GRB afterglow spectra (Christensen et al.\ 2011). The redshift interval is from $z=0.9$ (blue end of the spectrum) to $z=1.5$ (red end of the spectrum). The metallicity is not measured in the sample, but cold-ISM absorption lines are strong, especially in optically dark bursts. Dark bursts are mostly not seen in the optical and have an optical to X-ray spectral index $\beta_{\rm OX} < 0.5$. The global full composite includes the spectra of 60 GRB afterglows, mainly in the redshift interval $2 < z  < 2.4$, with an average metallicity which is relatively high: $Z/Z_\odot \sim 1/6$. In the same figure, we see the comparison with the composite spectrum of gas-rich QSO-DLAs at $2 < z < 4$, characterized by $N_{\rm HI} > 5\times10^{21.0}$ cm$^{-2}$ and relatively low metallicity, with $Z/Z_\odot \sim 1/20$ (Noterdaeme et al.\,2014).

\begin{figure*}\label{f7}
\epsfysize=5.8cm
\centerline{\epsfbox{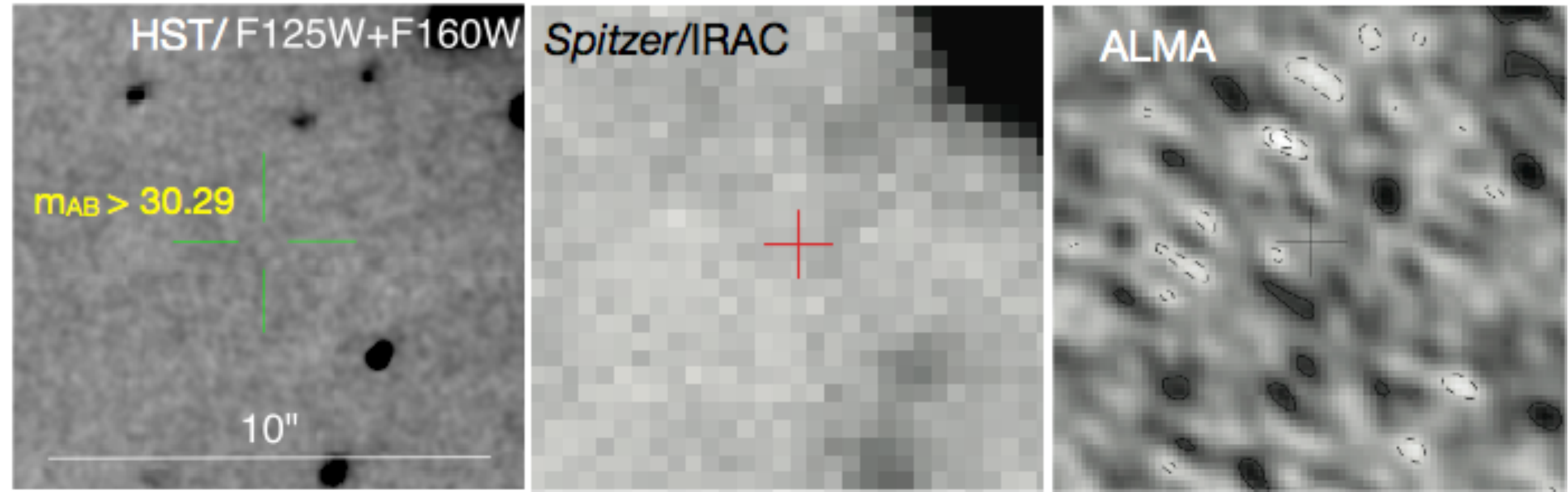}}
\caption{Observations of the field of GRB 090423 ($z=8.23$). {\it Left, center and right}: HST (Tanvir et al.\,2012), {\it Spitzer} (Berger et a.\,2014) and ALMA (Berger et al.\,2014) images. The scale is the same for all images. Ten arcsecs at $z=8.23$ corresponds to a physical size of 48 kpc. The position of the GRB (indicated by the cross and obtained from NIR imaging of the afterglow; Tanvir et al.\,2009b) is known with a precision of 0.3 arcsecs (1.45 kpc).}
\end{figure*}

A detailed comparison with more composite spectra, representing a large variety of galaxy populations, is given in Savaglio (2013). The average spectrum of 13 massive galaxies at $z \sim 1.6$ (median stellar mass $M_\ast = 2.4\times10^{10}$ M$_\odot$, SFR $= 30$ M$_\odot$\,yr$^{-1}$, specific star-formation rate $s$SFR $= 1.2$ Gyr$^{-1}$; Savaglio et al.\ 2004) shows similar absorptions to the GRB composite in Fig.\,5. The composite of a complete sample of UV-bright $z \sim1$ galaxies has much weaker  absorption lines, but sizable emission lines (from the hot gas), with the tendency of stronger absorbers to be more common in brighter galaxies (Martin et al.\ 2012). These galaxies have ${\rm SFR} =1-100$ M$_\odot$\,yr$^{-1}$ and $M_\ast = 10^{9.5}-10^{11.3}$ M$_\odot$\,yr$^{-1}$ ($s$SFR $ = 0.07 - 6$ Gyr$^{-1}$). A similar composite is the one of 28 local ($z<0.05$) starburst and star-forming galaxies, with median metallicity $\log Z/Z_\odot = -0.5$, UV luminosity and $K$-band absolute magnitude $L_{1500} = 5\times 10^{39}$ erg\,s$^{-1}$\,\AA$^{-1}$ and $M_K = -21.35$, respectively
(Leitherer et al.\ 2011). Using the empirical relations in Savaglio et al.\ (2009), these numbers are translated into a star-formation rate SFR$_{1500} \sim 1$ M$_\odot$\,yr$^{-1}$ and a stellar mass $M_\ast \sim 6\times10^{9}$ M$_\odot$ (assuming $A_V= 1$). This gives
a (rather uncertain) $s$SFR of a few Gyr$^{-1}$. Surprisingly, these values are not very dissimilar from those of the $z\sim0.75$ GRB host sample (Savaglio et al.\ 2009), despite the apparent difference with the GRB composite. However, we notice that the median redshift of the GRB host sample is lower than the redshift interval covered by the GRB composite ($z=0.9-1.5$), indicating again a redshift evolution of the galaxy population hosting GRBs.

\section{More massive GRB hosts at $z>1.5$}

As described above, at $z>1.5$, some GRB hosts are metal rich, massive, dusty (dark GRBs), or highly star forming (Hunt et al.\ 2011; Kr\"uhler et al. 2011; Rossi et al.\ 2012; Hunt et al.\,2014).  This kind of galaxies can be very bright in the sub-millimeter. A few examples, detected with Herschel and ALMA, are shown in Fig.\,6 (Hunt et al.\,2014; Hatsukade et al\,2014). However, a systematic search in a large sample at $z<1$ shows that the total number of radio bright hosts is very small (Micha{\l}owski et al.\,2012; Perley \& Perley 2013). One possible explanation is that sub-mm galaxies (SMGs) account for at most 20\% of the cosmic SFRD (Micha{\l}owski et al.\ 2010).  Future surveys can bridge the gap at $2 < z < 4$ and explain a possible steep redshift evolution. 

The fraction of pair absorbers in $z>1.5$ GRB afterglow spectra has been found to be almost three times higher than in QSO-DLAs (which probe random galaxies), suggesting that galaxy interactions may play a role in the formation of massive stars at high redshift (Savaglio et al.\ 2012). Another indication is the large fraction (at least 40\%) of known GRB hosts at $z > 1.5$ showing interaction, disturbed morphologies, or galaxy pairs (Th\"one et al.\ 2011; Vergani et al.\  2011; Chen 2012; Kr\"uhler et al.\ 2012). The interaction hypothesis is not surprising if one considers the higher fraction of galaxy mergers seen in the past of the Universe with respect to today (Bluck et al.\ 2012).

\section{High-redshift GRB hosts}

The investigation of GRB hosts at $z>4$ has been so far particularly difficult. At $z>4$ and $z>5$, we know 27 and 10 GRBs with measured redshift, respectively. Intensive search of $4<z <5$ GRB fields resulted in rest-frame UV (e.g., star formation) detections for 7 hosts. Detailed analysis of the luminosity function for detected (and non-detected) hosts at $3<z<5$ suggests that these galaxies might represent the most known population of galaxies at those redshifts, the Lyman break galaxies (Greiner et al.\,2015). However, at $z\geq5$, 5 GRB fields have been observed (Basa et al.\ 2012; Tanvir et al.\ 2012, Berger et al.\,2014), with negative results. The lack of Lyman break galaxies at very high redshift is also found by Schulze et al.\,(2015), though results are not totally consistent with those in Greiner et al.\,(2015. If no dust correction is assumed (dust is not expected to be abundant in a  less than 1.2 Gyr old Universe), the UV-luminosity limit $L_{1500}$ can be translated into a SFR limit (Savaglio et al.\ 2009). The deepest search obtained with HST/NIR  (Tanvir et al.\,2012), Spitzer medium IR,  and ALMA radio observations (Berger et al.\,2014) for the host of GRB\,090423 at $z=8.23$ gave negative results (Fig.\,7). The limiting magnitude obtained with HST in the NIR is $m_{\rm AB} > 30.29$, which corresponds to a very low upper limit for the UV luminosity, $L_{1500} < 3.7\times10^{38}$ erg s$^{-1}$, and an exceptionally low SFR $ < 0.07$ M$_\odot$\,yr$^{-1}$.

We can compare these low SFR limits to the stellar mass expected from numerical simulations. About 70\% of the hosts at $z>6$ predicted by Salvaterra et al.\ (2013) have stellar mass  in the range $M_\ast = 10^6-10^8$ M$_\odot$, while the star-formation rate and metallicity are in the intervals SFR $=0.03-0.3$ M$_\odot$\,yr$^{-1}$ and $Z/Z_\odot = 0.01-0.1$, respectively. The SFR limit for the host of GRB\,090423 indicates that a very low stellar mass, $M_\ast \sim 10^6$ M$_\odot$, is possible if $s$SFR $<70$ Gyr$^{-1}$. Vice versa, if we assume $s$SFR $\sim 10$ Gyr$^{-1}$, the limit SFR $ < 0.07$ M$_\odot$\,yr$^{-1}$ gives $M_\ast < 7\times10^6$ M$_\odot$. We conclude that in the past, GRB hosts must have been very small. 

\section{Are some high-$z$ GRBs hosted by young star clusters?}

\begin{figure}\label{f8}
\epsfysize=6cm
\centerline{\epsfbox{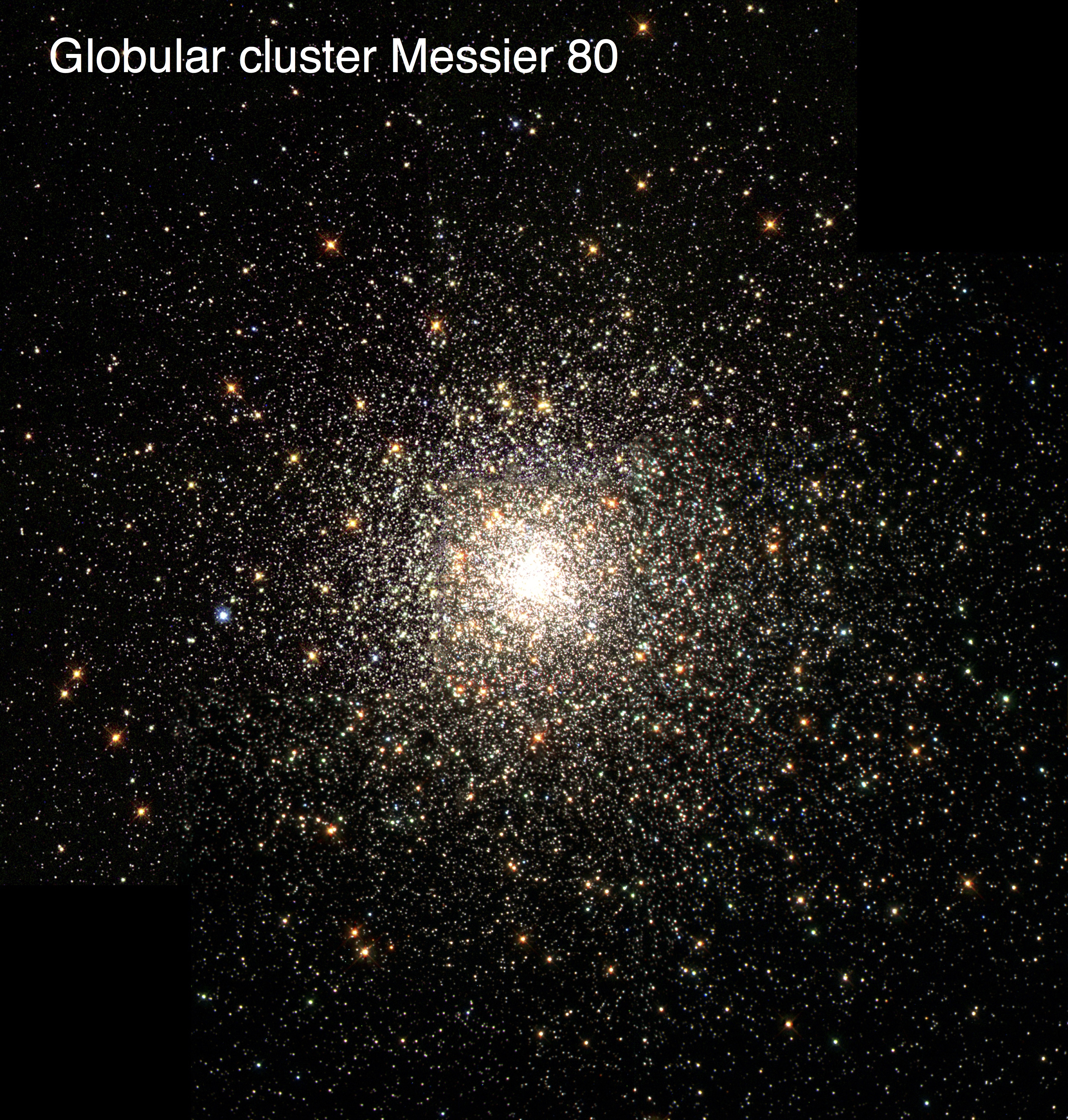}}
\epsfysize=5.8cm
\centerline{\epsfbox{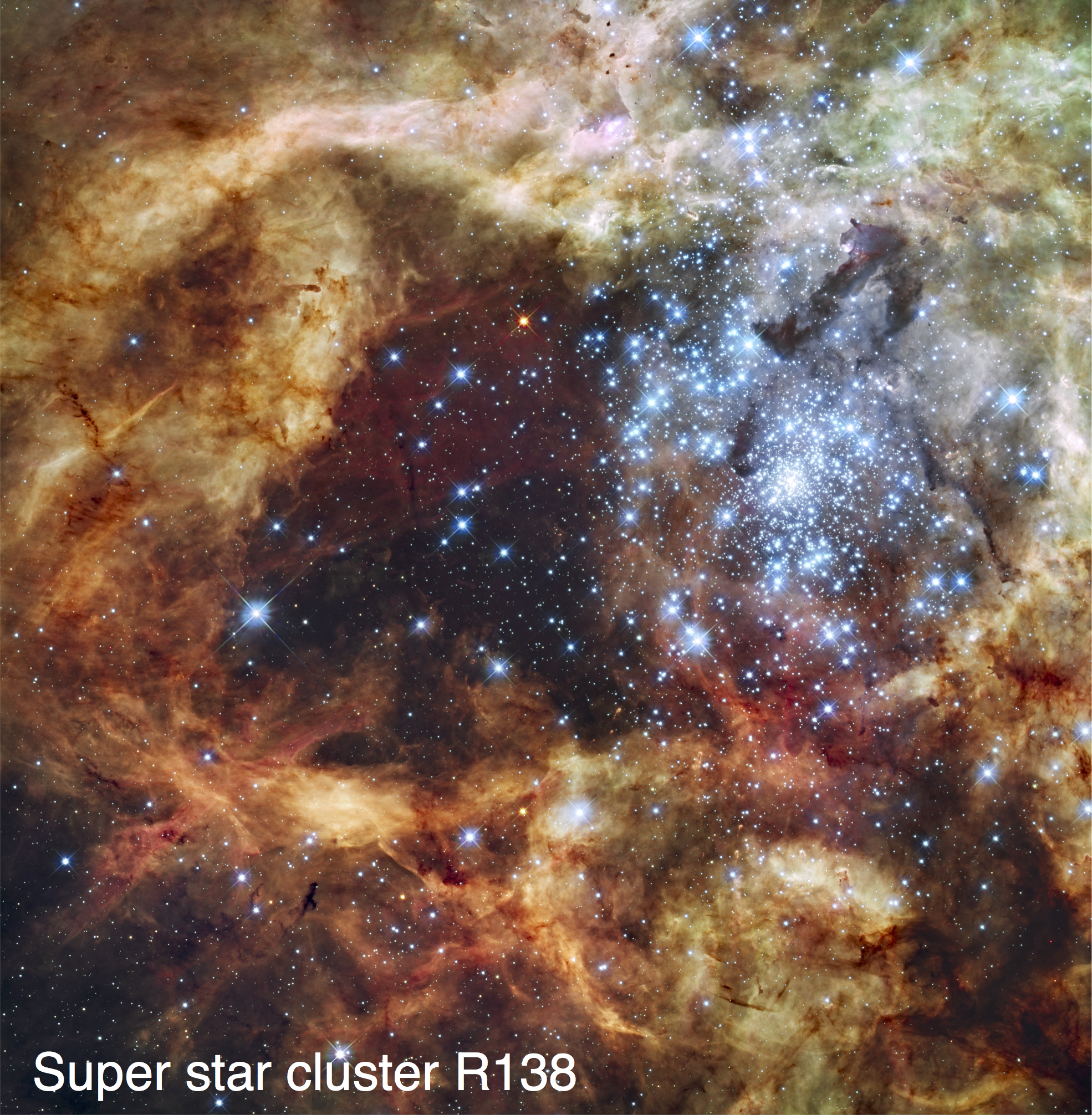}}
\caption{{\it Top}: HST/WFPC2 image of one of the densest globular clusters (GCs) known in the Milky Way: Messier 80 (credit: The Hubble Heritage Team -- AURA/ STScI/ NASA). The estimated mass, metallicity, and age are $M \sim 5\times 10^5$ M$_\odot$, $\log Z/Z_\odot = -1.47$, and $t \sim 12.5$ Gyr, respectively. Its age corresponds to a redshift of formation $z\sim5$. At that time, this GC must have been dominated by young and massive stars, with many of them being in binary or multiple systems. {\it Bottom}: HST UV, V and R-band image of the super star cluster R136, in the region 30 Doradus of the LMC (credit: R.\,O'Connell, University of Virginia, and the WFC3 Science Oversight Committee). Similar massive star clusters are the progenitors of GCs, and might host a GFB during the present active phase.}
\end{figure}

How about if a fraction of GRBs at high redshift happened in young progenitors of today's globular clusters? If so, finding a host like that might still be out of present telescope capabilities. The upper panel of Fig.\,\ref{f8} shows a beautiful HST image of a well known globular cluster (GC): Messier 80 (M80). Due to its intrinsic brightness (absolute magnitude $\sim - 7$) and proximity to the Sun (10 kpc), this is one of the most studied GCs ever. M80 is characterized by the usual properties for GCs: old (12.5 Gyr), metal poor ($\log Z/Z_\odot \simeq -1.5$), compact, with a high stellar density, and is also relatively massive ($M \simeq 5\times10^5$ M$_\odot$). If we believe that the Universe today is 13.7 Gyr old, M80 must have formed when the Universe was 1.2 Gyr old. That means that GRBs at $z\simeq5$ would be contemporary to the newly born M80. However, one main  problem of a proto-GC as a possible host of high-$z$ GRBs is the small fraction of stellar mass contained in GCs with respect to the total stellar mass in galaxies. Less than 200 are the known GCs in the MW.  If we assume they all have a mass similar to M80 (which is a massive GC) and that the stellar mass of the MW is $M_\ast(MW) \simeq 4\times10^{10}$ M$_\odot$, the fraction of stellar mass in GCs with respect to the total cannot be higher than $M_\ast(\rm GCs)/M_\ast(\rm MW) \sim 0.25$\%, very low. 

However, the picture cannot be discarded too quickly. Young proto-GCs must have had very different properties, perhaps similar to a young massive star cluster today. Moreover, it has been suggested that intense star formation in starbursts create the physical conditions to form  numerous and massive star clusters (e.g., Adamo et al.\,2011). Finally, hydrodynamical simulations have shown that galaxy collision has a role in the formation of massive star clusters (Renaud et al.\,2015), which eventually will evolve in GCs. As mentioned in Section 6, interactions and mergers are common in high-$z$ GRBs. 

The lower panel of Fig.\,\ref{f8} shows the recently discovered massive star cluster R136, in the star-forming region 30 Doradus of the LMC (Crowthler et al.\,2010). This star cluster is young (a few Myr old), and as massive as M80. What is interesting from the point of view of GRBs is that R136 contains a high concentration of massive stars, over 70 O and Wolfe-Rayet stars within 5 pc.
Equally important, Conroy (2012) found indications that the initial mass of GCs were at least $10-20$ times larger than today, easily exceeding $10^6$ M$_\odot$. Mass loss is due to stellar winds, shocks and SN events. Moreover, a certain number of GCs in the MW might have disappeared early on, being gravitationally unbound within 1 Gyr (Vesperini et al.\,2009), or disrupted during the merging of proto-galaxies.

From the Galaxy point of view, it has been known for a decade now that a large fraction of the cosmic stellar mass was produced at $z >1$, when the Universe was almost half of its present age. In a recent review article, Madau \& Dickinson (2014) show that the stellar-mass density from $z=5$ to now has increased by a factor of $\sim60$  (Fig.\,9). A galaxy like the MW was still in the process of forming at $z=5$, very likely the main body of its mass (for instance, the disk) was not in place jet, whereas likely a proto-bulge was. However, a large fraction of the proto-GCs, those that today are older than 12.5 Gyr were already formed or on the process of forming, some will never survive for long, suffering from the disruptive gravitational force of the MW. 

All together, this means that the initial fraction of stars in proto-GCs must have been much larger than the 0.25\% fraction of today. We quantify under which circumstances, at $z=5$, the total mass in proto-GCs would be about the mass of the proto-MW. If we assume that: $(a)$ proto-GCs were 10 times more massive, which means $M($proto-GC$) \sim5\times10^6$ M$_\odot$ $(b)$ the total number of proto-GCs was lower than today because many formed at $z < 5$; $(c)$ but this is compensated by assuming that many first generation proto-GCs did not survive until today; $(d)$ the stellar mass of the MW was 40 times lower, e.g., $M($proto-MW$)\sim 1\times10^{10}$ M$_\odot$, then all these conditions together give a ratio of stellar mass in proto-GCs to the one of the proto-MW $M_\ast($proto-GCs$)/M_\ast($proto-MW$) \sim 1$. Stars in proto-GCs and the proto-MW are equally important.

Another last situation in favor of proto-GCs as hosts of high-$z$ GRBs is that they have a high density of binary systems. Massive and rapidly rotating binary systems are expected to be very common in a low-metallicity environment (Sana et al., 2012) and such a configuration is considered to be very favorable for high-redshift GRBs (Bromm \& Loeb(2006). For the description of the formation of fast-rotating massive stars in a possible scenario of proto-GCs, see Krause et al.\,(2013).

In Fig.\,10, we show this possible scenario, with a GRB happening in a proto-GC at $z=5$. On top an optical image of the Andromeda galaxy. In the center a UV image of the same galaxy, with artificially overlaid many GCs like M80. At the bottom, we imagine that at $z=5$, only the bulge of such a giant spiral galaxy was in place, the UV part in the disk with young stars today would form later, while all GCs were already there. In the same image, a GRB has exploded inside a GC.

\begin{figure}\label{f9}
\epsfysize=7cm
\centerline{\epsfbox{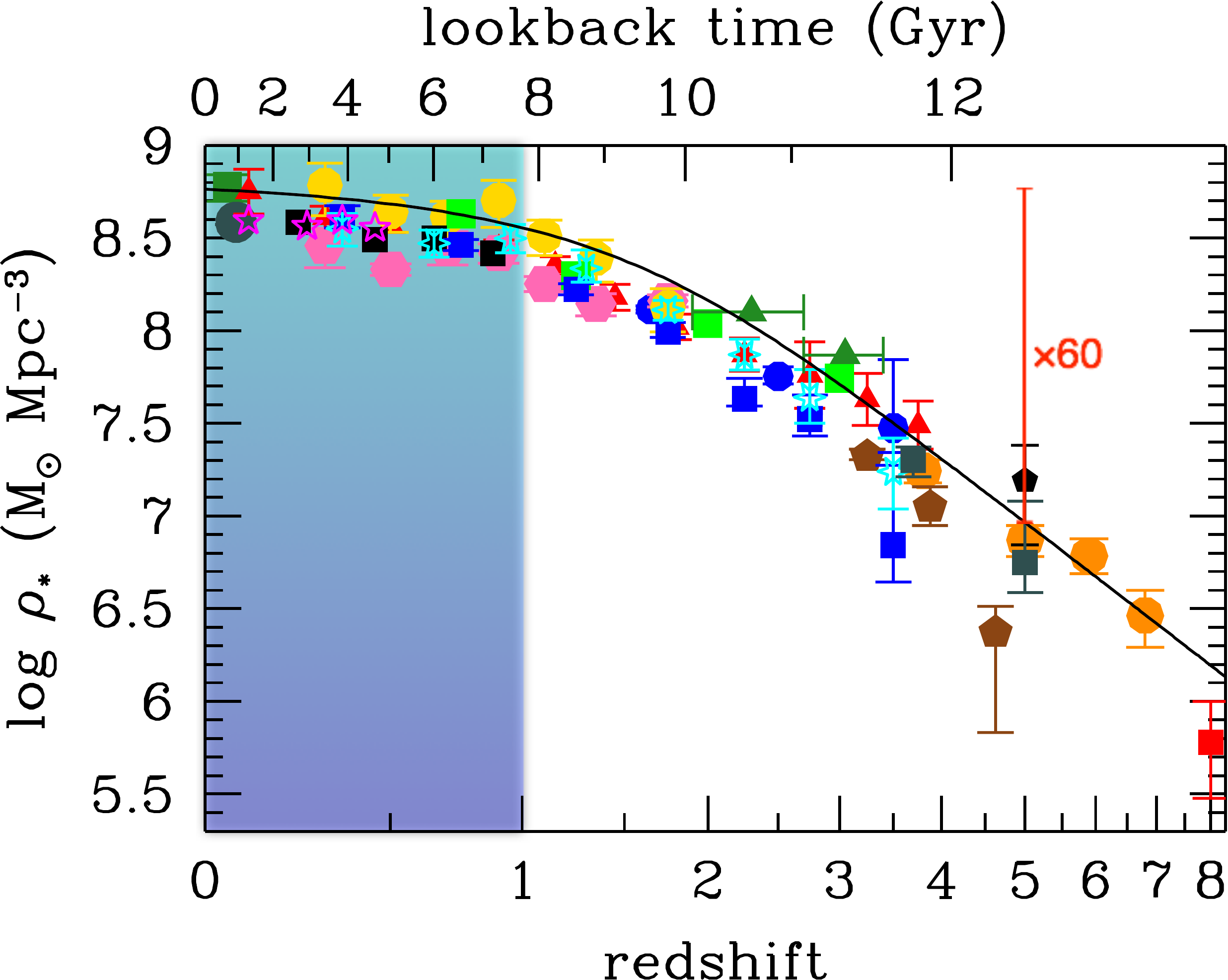}}
\caption{The evolution of the stellar mass density in the Universe (from Madau \& Dickinson 2014). The data points are collected from papers published since 2007. The solid line shows the prediction generated by integrating the instantaneous star formation rate density as a function of redshift. The increase in the cosmic stellar mass from $z=5$ to today was a factor of $\sim60$. Most of the action took place at $z>1$. Below this redshift (shaded area), the Universe (which was less than 6 Gyr old, or 43\% of its present age) had already formed more than 60\% of the stars today.}
\end{figure}

\begin{figure}\label{f10}
\epsfysize=5.5cm
\centerline{\epsfbox{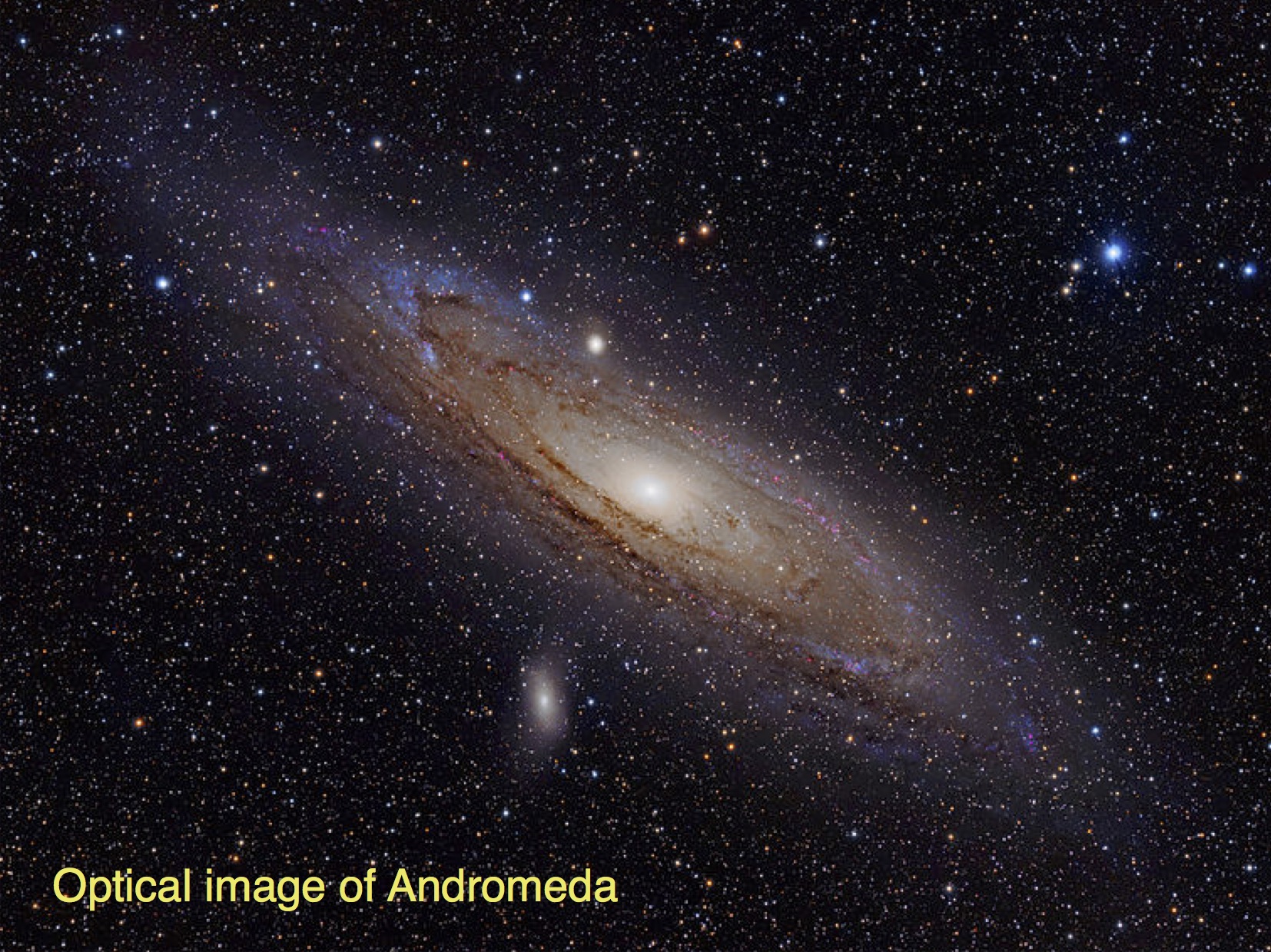}}
\centerline{\epsfbox{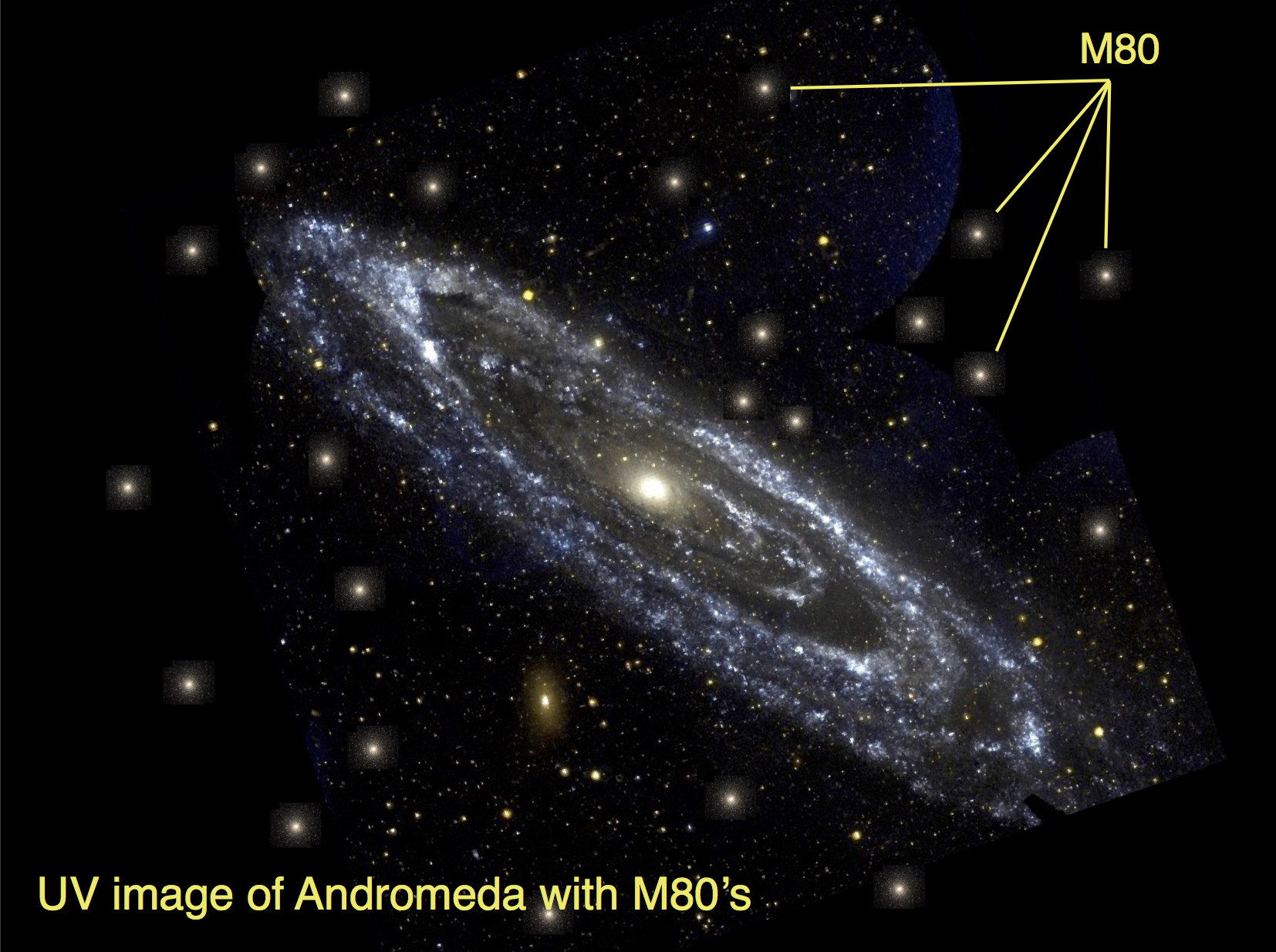}}
\centerline{\epsfbox{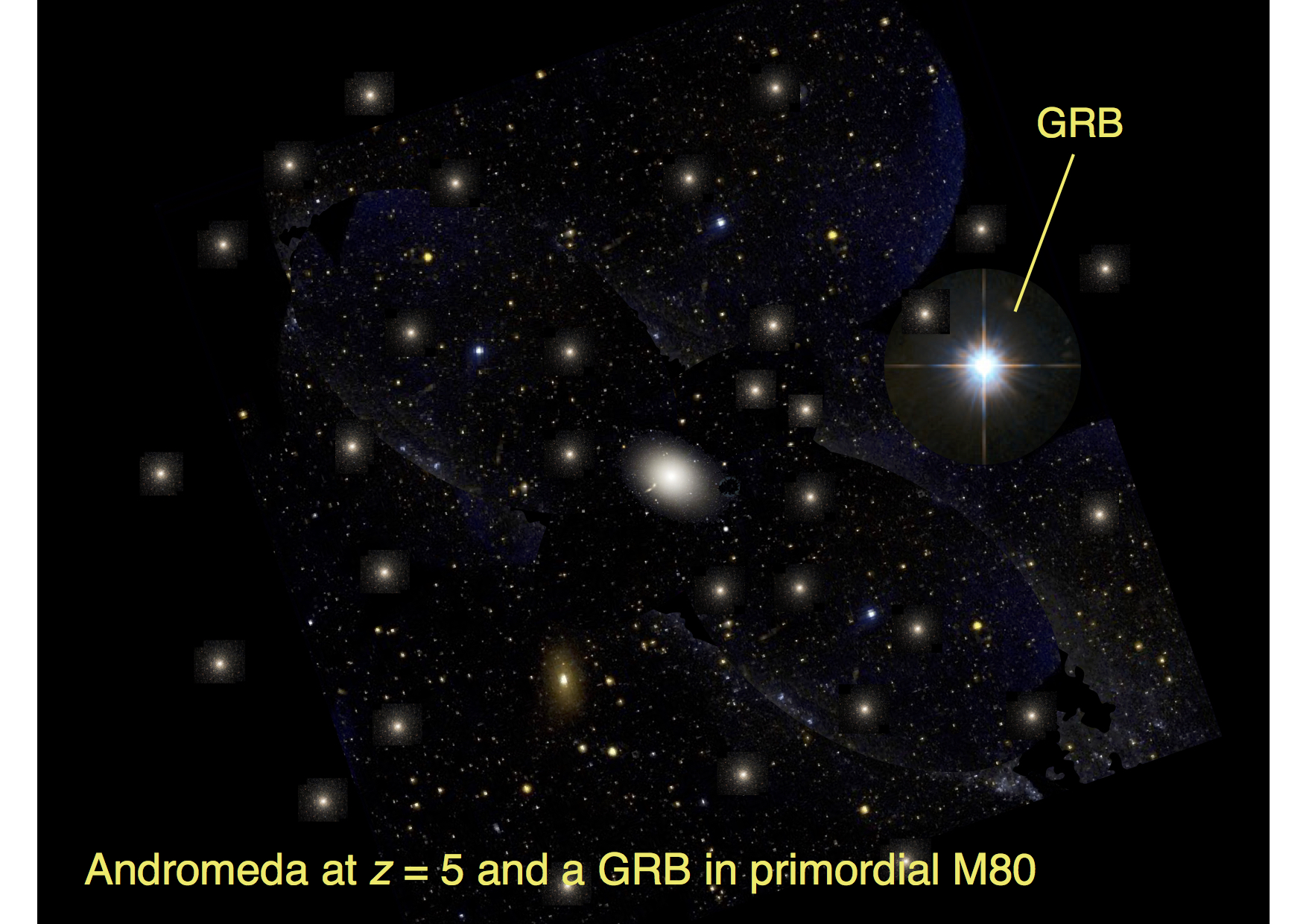}}
\caption{A possible scenario for the origin of $z>5$ GRBs. {\it Top}:  optical image of Andromeda; {\it at center}:  UV image of the same galaxy, with artificially overlaid several  globular clusters as M80; {\it bottom}: the imaginary proto-bulge of the same galaxy, no disk present yet, and the same much younger proto-GCs. In one of the them, the explosion of a massive star in a binary system became a GRB.}
\end{figure}

\section{Conclusions}

The impact of GRB host galaxies on the understanding of galaxy formation and evolution is still affected by small number statistics. Their knowledge is mainly limited to the $z<1.5$ regime, where about half of the galaxies hosting GRBs have been found and studied in detail. From this, the most accepted picture is that GRB hosts are generally small, star forming and metal and dust poor objects. However, at $z>1.5$, metallicity, mass and dust extinction show a large spread, suggesting a different population. Many host galaxies show disturbed morphologies, interactions with nearby galaxies and mergers. All this is nicely connected to the idea that local massive ellipticals today were young and bursty in the past, with some of them experiencing close encounters with other galaxies, which likely triggered intense episodes of star formation. 

At very high redshift, $z>5$, the situation might have changed again. Massive galaxies were very rare, but these are observationally easier to find. In fact, at these distances, deep searches failed to detected any GRB host, and relatively low SFRs were inferred. Unless dust content was already important back then (unlikely), low SFRs means low galaxy mass. Therefore, GRB hosts in the past could have been more similar to the local counterparts. This is supported by recent findings according to which the UV flux that ionized the Universe at high redshift was produced by very small, young and star forming galaxies (Bouwens et al.\,2015). 

As an alternative picture, we propose a different paradigm, never considered in detail, according to which a proto globular cluster, which would look like a massive young star cluster in the local Universe, could host high-$z$ GRBs. Globular clusters today contain a very low fraction of the total stellar mass, but at $z>5$ proto-GCs can be equally important in terms of stellar mass because many lost a large fraction of the mass in subsequent encounters with more massive galaxies. Moreover, giant galaxies like the MW have formed a large fraction of their present stellar mass in recent times. In support of this idea is the fact that proto-GCs probably contains a large fraction of massive and rapidly rotating binary systems, also a favorable situation for GRBs.

In summary, GRB hosts $z <1.5$ are generally small galaxies, whereas at intermediate redshift an important fraction of hosts are massive, dusty or metal rich galaxies. Going further in the past, in the primordial Universe, host galaxies are again likely very small systems, so small that, perhaps, these systems are massive young star clusters with a large fraction of massive and rotating binary systems, possible seeds of GRBs.

\section*{Acknowledgments}

I wish to thank the organizers for the kind invitation, and the referee for comments and suggestions which stimulated a more detailed discussion about GCs as possible high-$z$ GRB hosts. I express my appreciation to Enzo Brocato for sharing his knowledge of globular clusters and the young star cluster R136.




\begin{thebibliography}{00}

\bibitem[Adamo et al.(2011)]{2011MNRAS.417.1904A} Adamo, A., {\"O}stlin, G., \& Zackrisson, E.\ 2011, MNRAS, 417, 1904

\bibitem[Basa et al.(2012)]{2012A&A...542A.103B} Basa, S., Cuby, J.~G., Savaglio, S., et al.\ 2012, A\&A, 542, A103 

\bibitem[Berger et al.(2014)]{2014ApJ...796...96B} Berger, E., Zauderer, B.~A., Chary, R.-R., et al.\ 2014, ApJ, 796, 96 

\bibitem[Bloom et al.(2002)]{bloom02} Bloom, J.~S., Kulkarni, S.~R., Price, P.~A., et al.\ 2002, ApJ, 572, L45

\bibitem{bluck} Bluck, A.\,F.\,L., Conselice, C.\,J., Buitrago, F., et al.\ 2012, ApJS, 747, 34

\bibitem[Boissier et  al.(2013)]{2013A&A...557A..34B} Boissier, S., Salvaterra, R., Le Floc'h, E., et al.\ 2013, A\&A, 557, A34 

\bibitem[Bouwens et al.(2015)]{2015arXiv150308228B} Bouwens, R.~J., Illingworth, G.~D., Oesch, P.~A., et al.\ 2015, arXiv:1503.08228 

\bibitem[Bromm \& Loeb(2006)]{2006ApJ...642..382B} Bromm, V., \& Loeb, A.\ 2006, ApJ, 642, 382 

\bibitem[Campisi et al.(2011)]{2011MNRAS.417.1013C} Campisi, M.~A., Tapparello, C., Salvaterra, R., Mannucci, F., \& Colpi, M.\ 2011, MNRAS, 417, 1013 

\bibitem{chen} Chen, H.-W.\ 2012, MNRAS, 419, 3039

\bibitem[Chary et al.(2007)]{2007ApJ...671..272C} Chary, R., Berger, E., \& Cowie, L.\ 2007, ApJ, 671, 272 

\bibitem[Chisari et al.(2010)]{2010MNRAS.408..647C} Chisari, N.~E., Tissera, P.~B., \& Pellizza, L.~J.\ 2010, MNRAS, 408, 647 

\bibitem[Christensen et al.(2011)]{2011ApJ...727...73C} Christensen, L., Fynbo, J.~P.~U., Prochaska, J.\,X., et al.\ 2011, ApJ, 727, 73 

\bibitem[Conroy(2012)]{2012ApJ...758...21C} Conroy, C.\ 2012, ApJ, 758, 21 

\bibitem[Costa et al.(1997)]{1997Natur.387..783C} Costa, E., Frontera, F., Heise, J., et al.\ 1997, Nature, 387, 783 

\bibitem[Crowther et al.(2010)]{2010MNRAS.408..731C} Crowther, P.~A., Schnurr, O., Hirschi, R., et al.\ 2010, MNRAS, 408, 731 

\bibitem[Djorgovski et al.(1999)]{1999GCN...289....1D} Djorgovski, S.~G., Kulkarni, S.~R., Bloom, J.~S., \& Frail, D.~A.\ 1999, GRB Coordinates Network, 289, 1 

\bibitem[Elliott et al.(2012)]{2012A&A...539A.113E} Elliott, J., Greiner, J., Khochfar, S., et al.\ 2012, A\&A, 539, A113 

\bibitem[Friis et al.(2014)]{2014arXiv1409.6315F} Friis, M., De Cia, A., Kr{\"u}hler, T., et al.\ 2014, arXiv:1409.6315 

\bibitem[Fruchter et al.(2006)]{2006Natur.441..463F} Fruchter, A.~S., Levan, A.~J., Strolger, L., et al.\ 2006, Nature, 441, 463 

\bibitem[Fynbo et al.(2006)]{2006A&A...451L..47F} Fynbo, J.~P.~U., Starling, R.~L.~C., Ledoux, C., et al.\ 2006, A\&A, 451, L47

\bibitem[Fynbo et al.(2014)]{2014A&A...572A..12F} Fynbo, J.~P.~U., Kr{\"u}hler, T., Leighly, K., et al.\ 2014, A\&A, 572, A12 

\bibitem[Galama et al.(1998)]{1998Natur.395..670G} Galama, T.~J., Vreeswijk, P.~M., van Paradijs, J., et al.\ 1998, Nature, 395, 670

\bibitem[Gehrels et al.(2004)]{2004ApJ...611.1005G} Gehrels, N., Chincarini, G., Giommi, P., et al.\ 2004, ApJ, 611, 1005 

\bibitem[Graham \& Fruchter(2013)]{2013ApJ...774..119G} Graham, J.~F., \& Fruchter, A.~S.\ 2013, ApJ, 774, 119 

\bibitem[Greiner et al.(2015)]{2015arXiv150305323G} Greiner, J., Fox, D.~B., Schady, P., et al.\ 2015, ApJ, submitted, arXiv:1503.05323 

\bibitem[Greiner et al.(2011)]{2011A&A...526A..30G} Greiner, J., Kr{\"u}hler, T., Klose, S., et al.\ 2011, A\&A, 526, AA30 

\bibitem[Guimar{\~a}es et al.(2012)]{2012AJ....143..147G} Guimar{\~a}es, R., Noterdaeme, P., Petitjean, P., et al.\ 2012, AJ, 143, 147 

\bibitem[Hatsukade et al.(2014)]{2014Natur.510..247H} Hatsukade, B., Ohta, K., Endo, A., et al.\ 2014, Nature, 510, 247 

\bibitem[Hjorth \& Bloom(2012)]{2012grbu.book..169H} Hjorth, J., \& Bloom, J.~S.\ 2012, Chapter 9 in ''Gamma-Ray Bursts'', Cambridge Astrophysics Series 51,  eds.~C.~Kouveliotou, R.~A.~M.~J.~Wijers and S.~Woosley, Cambridge University Press (Cambridge), p.~169-190, 169 
\bibitem[Hjorth et al.(2012)]{2012ApJ...756..187H} Hjorth, J., Malesani, D., Jakobsson, P., et al.\ 2012, ApJ, 756, 187 

\bibitem[Hopkins \& Beacom(2006)]{2006ApJ...651..142H} Hopkins, A.~M., \& Beacom, J.~F.\ 2006, ApJ, 651, 142 

\bibitem{hunt} Hunt, L., Palazzi, E., Rossi, A., et al., 2011, ApJ, 736, L36

\bibitem[Hunt et al.(2014)]{2014A&A...565A.112H} Hunt, L.~K., Palazzi, E., Micha{\l}owski, M.~J., et al.\ 2014, A\&A, 565, AA112

\bibitem[Kann et al.(2006)]{2006ApJ...641..993K} Kann, D.~A., Klose, S., \& Zeh, A.\ 2006, ApJ, 641, 993 

\bibitem[Kann et al.(2010)]{2010ApJ...720.1513K} Kann, D.~A., Klose, S., Zhang, B., et al.\ 2010, ApJ, 720, 1513 

\bibitem[Kann et al.(2011)]{2011ApJ...734...96K} Kann, D.~A., Klose, S., Zhang, B., et al.\ 2011, ApJ, 734, 96 

\bibitem[Kelly et al.(2008)]{2008ApJ...687.1201K} Kelly, P.~L., Kirshner, R.~P., \& Pahre, M.\ 2008, ApJ, 687, 1201

\bibitem[Krause et  al.(2013)]{2013A&A...552A.121K} Krause, M., Charbonnel, C., Decressin, T., Meynet, G., \& Prantzos, N.\ 2013, A\&A, 552, A121 

\bibitem{kruhler2}  Kr{\"u}hler, T.,  Greiner, J., Schady, P., et al., 2011, A\&A, 534, A108

\bibitem[Kr{\"u}hler et al.(2012)]{2012A&A...546A...8K} Kr{\"u}hler, T., Fynbo, J.~P.~U., Geier, S., et al.\ 2012, A\&A, 546, A8

\bibitem[Kr{\"u}hler et al.(2015)]{2015arXiv150506743K} Kr{\"u}hler, T., Malesani, D., Fynbo, J.~P.~U., et al.\ 2015, A\&A in press, arXiv:1505.06743

\bibitem[Kr{\"u}hler et al.(2008)]{2008ApJ...685..376K} Kr{\"u}hler, T., K{\"u}pc{\"u} Yolda{\c s}, A., Greiner, J., et al.\ 2008, ApJ, 685, 376 

\bibitem[Kr{\"u}hler et al.(2013)]{2013A&A...557A..18K} Kr{\"u}hler, T., Ledoux, C., Fynbo, J.~P.~U., et al.\ 2013, A\&A, 557, AA18 

\bibitem[K{\"u}pc{\"u} Yolda{\c s} et al.(2007)]{2007A&A...463..893K} K{\"u}pc{\"u} Yolda{\c s}, A., Salvato, M., Greiner, J., et al.\ 2007, A\&A, 463, 893 

\bibitem[Leitherer et al.(2011)]{2011AJ....141...37L} Leitherer, C., Tremonti, C.~A., Heckman, T.~M., et al.\ 2011, AJ, 141, 37

\bibitem[Ledoux et al.(2015)]{2015arXiv150407254L} Ledoux, C., Noterdaeme, P., Petitjean, P., \& Srianand, R.\ 2015, A\&A, in press, arXiv:1504.07254 

\bibitem{levesque} Levesque, E.,  Kewley, L.\,J., Berger, E., et al.\ 2010a, AJ, 140, 1557

\bibitem[Levesque et al.(2010)]{2010ApJ...712L..26L} Levesque, E.~M., Kewley, L.~J., Graham, J.~F., \& Fruchter, A.~S.\ 2010b, ApJL, 712, L26 

\bibitem[Li(2008)]{2008MNRAS.388.1487L} Li, L.-X.\ 2008, MNRAS, 388, 1487

\bibitem[Madau \& Dickinson(2014)]{2014ARA&A..52..415M} Madau, P., \& Dickinson, M.\ 2014, ARA\&A, 52, 415 

\bibitem[Martin et al.(2012)]{2012ApJ...760..127M} Martin, C.~L., Shapley, A.~E., Coil, A.~L., et al.\ 2012, ApJ, 760, 127 

\bibitem[Metzger et al.(1997)]{1997Natur.387..878M} Metzger, M.~R., Djorgovski, S.~G., Kulkarni, S.~R., et al.\ 1997, Nature, 387, 878

\bibitem{michalowski3} Micha{\l}owski, M., Hjorth, J., \& Watson, D.\ 2010, A\&A, 514, A67

\bibitem{michalowski2} Micha{\l}owski, M.~J., Kamble, A., Hjorth, J., et al.\ 2012, ApJ, 755, 85

\bibitem[Muzahid et al.(2015)]{2015MNRAS.448.2840M} Muzahid, S., Srianand, R., \& Charlton, J.\ 2015, MNRAS, 448, 2840 

\bibitem[Niino et al.(2011)]{2011ApJ...726...88N} Niino, Y., Choi, J.-H., Kobayashi, M.~A.~R., et al.\ 2011, ApJ, 726, 88

\bibitem[Niino et al.(2012)]{2012arXiv1204.0583N} Niino, Y., Hashimoto, T., Aoki, K., et al.\ 2012, PASJ, 64, 115

\bibitem[Noterdaeme et al.(2008)]{noterdaeme08} Noterdaeme, P., Ledoux, C., Petitjean, P., \& Srianand, R.\ 2008, A\&A, 481, 327 

\bibitem[Noterdaeme et al.(2014)]{2014A&A...566A..24N} Noterdaeme, P., Petitjean, P., P{\^a}ris, I., et al.\ 2014, A\&A, 566, AA24 

\bibitem[Perley et al.(2015)]{2015arXiv150402482P} Perley, D.~A.,  Kr{\"u}hler, T., Schulze, S., et al.\ 2015a, ApJ, submitted, arXiv:1504.02482 

\bibitem[Perley et al.(2013)]{2013ApJ...778..128P} Perley, D.~A., Levan, A.~J., Tanvir, N.~R., et al.\ 2013, ApJ, 778, 128 

\bibitem[Perley et al.(2012)]{2012ApJ...758..122P} Perley, D.~A., Modjaz, M., Morgan, A.~N., et al.\ 2012, ApJ, 758, 122 

\bibitem[Perley \& Perley(2013)]{2013ApJ...778..172P} Perley, D.~A., \& Perley, R.~A.\ 2013, ApJ, 778, 172 

\bibitem[Perley et al.(2015)]{2015arXiv150402479P} Perley, D.~A., Tanvir, N.~R., Hjorth, J., et al.\ 2015b, ApJ, submitted, arXiv:1504.02479 

\bibitem[Prochaska et al.(2009)]{2009ApJ...691L..27P} Prochaska, J.~X., Sheffer, Y., Perley, D.~A., et al.\ 2009, ApJ, 691, L27

\bibitem[Rau et al.(2010)]{2010ApJ...720..862R} Rau, A., Savaglio, S., Kr{\"u}hler, T., et al.\ 2010, ApJ, 720, 862 

\bibitem[Renaud et al.(2015)]{2015MNRAS.446.2038R} Renaud, F., Bournaud, F., \& Duc, P.-A.\ 2015, MNRAS, 446, 2038 

\bibitem{rossi} Rossi, A.,  Klose, S., Ferrero, P., et al.\ 2012, A\&A, 545, A77

\bibitem[Salvaterra et al.(2009)]{2009Natur.461.1258S} Salvaterra, R., Della Valle, M., Campana, S., et al.\ 2009, Nature, 461, 1258 

\bibitem[Salvaterra et al.(2013)]{2013MNRAS.429.2718S} Salvaterra, R., Maio, U., Ciardi, B., \& Campisi, M.~A.\ 2013, MNRAS, 429, 2718

\bibitem[Sana et al.(2012)]{2012Sci...337..444S} Sana, H., de Mink, S.~E., de Koter, A., et al.\ 2012, Science, 337, 444

\bibitem[]{savage77} Savage, B.\,D., Bohlin, R.\,C., Drake, J.\,F., Budich, W., 1977, ApJ, 216, 291

\bibitem[Savaglio(2006)]{2006NJPh....8..195S} Savaglio, S.\ 2006, New Journal of Physics, 8, 195 

\bibitem[Savaglio(2013)]{2013EAS....61..381S} Savaglio, S.\ 2013, EAS Publications Series, 61, 381 

\bibitem[Savaglio et al.(2004)]{2004ApJ...602...51S} Savaglio, S., Glazebrook, K., Abraham, R.~G., et al.\ 2004, ApJ, 602, 51 

\bibitem[Savaglio et al.(2009)]{2009ApJ...691..182S} Savaglio, S., Glazebrook, K., \& Le Borgne, D.\ 2009, ApJ, 691, 182

\bibitem{savaglio12} Savaglio, S., Rau, A.,  Greiner, J., et al.\,2012, MNRAS, 420, 627

\bibitem[Schady et al.(2011)]{2011A&A...525A.113S} Schady, P., Savaglio, S., Kr{\"u}hler, T., Greiner, J., \& Rau, A.\ 2011, A\&A, 525, AA113 

\bibitem[Schulze et al.(2015)]{2015arXiv150304246S} Schulze, S., Chapman, R., Hjorth, J., et al.\ 2015, ApJ, submitted, arXiv:1503.04246 

\bibitem[Tanvir et al.(2009)]{2009Natur.461.1254T} Tanvir, N.~R., Fox, D.~B., Levan, A.~J., et al.\ 2009a, Nature, 461, 1254 

\bibitem[Tanvir et al.(2009)]{2009GCN..9202....1T} Tanvir, N., Levan, A., Kerr, T., \& Wold, T.\ 2009b, GRB Coordinates Network, 9202

\bibitem[Tanvir et al.(2012)]{2012ApJ...754...46T} Tanvir, N.~R., Levan, A.~J., Fruchter, A.~S., et al.\ 2012, ApJ, 754, 46

\bibitem{thoene} Th{\"o}ne, C.,  Campana, S., Lazzati, D., et al.\ 2011, MNRAS, 414, 479

\bibitem[Trenti et al.(2015)]{2015ApJ...802..103T} Trenti, M., Perna, R., \& Jimenez, R.\ 2015, ApJ, 802, 103 

\bibitem[Tumlinson et al.(2007)]{2007ApJ...668..667T} Tumlinson, J., Prochaska, J.~X., Chen, H.-W., Dessauges-Zavadsky, M., \& Bloom, J.~S.\ 2007, ApJ, 668, 667

\bibitem[Tumlinson et al.(2002)]{tumlinson02} Tumlinson, J., Shull, J.~M., Rachford, B.~L., et al.\ 2002, ApJ, 566, 857

\bibitem[Vergani et al.(2011)]{2011AN....332..292V} Vergani, S.~D., Piranomonte, S., Petitjean, P., et al.\ 2011, Astron.\, Nach., 332, 292 

\bibitem[Vergani et al.(2014)]{2014arXiv1409.7064V} Vergani, S.~D., Salvaterra, R., Japelj, J., et al.\ 2014, A\&A,arXiv:1409.7064 

\bibitem[Vesperini et al.(2009)]{2009ApJ...698..615V} Vesperini, E., McMillan, S.~L.~W., \& Portegies Zwart, S.\ 2009, ApJ, 698, 615

\bibitem[Wang \& Dai(2009)]{2009MNRAS.400L..10W} Wang, F.~Y., \& Dai, Z.~G.\ 2009, MNRAS, 400, L10 

\bibitem[Wei et al.(2014)]{2014MNRAS.439.3329W} Wei, J.-J., Wu, X.-F., Melia, F., Wei, D.-M., \& Feng, L.-L.\ 2014, MNRAS, 439, 3329

\bibitem[Y{\"u}ksel et al.(2008)]{2008ApJ...683L...5Y} Y{\"u}ksel, H., Kistler, M.~D., Beacom, J.~F., \& Hopkins, A.~M.\ 2008, ApJ, 683, L5 

\bibitem[Zafar et al.(2012)]{2012ApJ...753...82Z} Zafar, T., Watson, D., El{\'{\i}}asd{\'o}ttir, {\'A}., et al.\ 2012, ApJ, 753, 82

\end{thebibliography}


\end{document}